\documentclass[twocolumn]{aastex631}
\usepackage{amsmath}
\usepackage{CJK}

\begin{document}

\begin{CJK*}{UTF8}{gbsn}
\title{Overall Ultra-high Energy Neutrino Emission from GRBs during Jet Expansion}

\correspondingauthor{Zhuo Li}
\email{zhuo.li@pku.edu.cn, zhangqy@stu.pku.edu.cn}

\author[0000-0003-1469-208X]{Qinyuan Zhang (张秦源)}
\affiliation{Department of Astronomy, School of Physics, Peking University, Beijing 100871, China}

\author{Zhuo Li (黎卓)}
\affiliation{Department of Astronomy, School of Physics, Peking University, Beijing 100871, China}
\affiliation{Kavli Institute for Astronomy and Astrophysics, Peking University, Beijing 100871, China}

\begin{abstract}
The ultra-relativistic jet released in gamma-ray bursts (GRBs) is expected to produce ultra-high-energy cosmic rays (UHECRs), prompt gamma-ray emission and hence prompt high-energy neutrinos by photopion interactions. In this work, we calculate the time-integrated neutrino spectrum during the expansion of jets by taking into account the time evolution of cosmic ray and secondary spectra and neutrino production. We numerically solve the continuity equations for nucleons, pions, and muons for their spectral evolution. Since pion and muon damping weakens as the jet expands, the neutrino production at large radii at high energies may dominate that around the jet energy dissipation radius. 
Compared with the usually adopted approaches that only consider neutrino production around the energy dissipation radius, the overall UHE neutrino fluence integrated over time can be significantly larger, and the flavor fraction of electron neutrinos as function of neutrino energy is different at UHE, due to neutrino production at radii much larger than the energy dissipation radius. The faster magnetic field decay leads to larger UHE neutrino fluence, and the UHE neutrino spectra is weakly dependent on the energy dissipation radius and the jet Lorentz factor. Observations of prompt EeV neutrinos from GRBs by the next-generation neutrino telescopes, e.g., GRAND and IceCube-Gen2, will test the hypothesis of GRBs as UHECR sources and probe the physics of GRB jets.
\end{abstract}



\section{Introduction} \label{sec:1}
\end{CJK*}
Gamma-ray bursts (GRBs) are the most violent and luminous events in the universe, releasing huge energy in seconds \citep[For review see, e.g.,][]{2019pgrb.book.....Z}. It is well established that the GRB central engine generates relativistic jets that produce prompt gamma-ray emission by internal energy dissipation \citep{1994ApJ...427..708P,1994ApJ...430L..93R}, and drive relativistic external shocks into a circumburst medium that produce long-lasting afterglow emission \citep{1993ApJ...418L...5P,1997ApJ...476..232M}.
GRBs are one of the main candidates for ultra-high energy cosmic ray (UHECR; $>10^{19}$eV) sources \citep[see review by, e.g.,][]{UHECR_ARAA_Kotera}. Their extreme physical condition, especially the large luminosity, implies that GRBs may accelerate nuclei up to $ \sim 10^{20} \rm eV$ \citep[e.g.,][]{GRBUHECR_Waxman1995,1995ApJ...453..883V}. Thus, GRBs are also expected to be high-energy neutrino sources, given that their intense prompt gamma-ray photons can naturally serve as targets for photopion interactions (PIs) \citep{Waxman_1997,Waxman_EeVafterglow,2006PhRvD..73f3002M}, which produces pions, and followed by neutrino generation from charged pion decay, $\pi^{\pm} \rightarrow \mu + \nu_{\mu}$, and hence muon decay $\mu \rightarrow e + \nu_{e} + \nu_{\mu}$.

Intense efforts had been carried out to search for neutrinos from GRBs \citep{2010ApJ...710..346A,2011PhRvL.106n1101A,2012Natur.484..351I}. However, so far no significant correlation between neutrino events and GRBs is found by IceCube \citep[see the latest results by][]{ICGRB2017,ICGRB2022}, leading to the conclusion that the prompt-phase neutrino emission from GRBs contributes $< 1\%$ of the observed diffuse neutrino flux \citep{ICGRB2022}. The non-detection of neutrinos, compared with correct modeling of GRB neutrino production \citep[e.g.,][]{2004APh....20..429G,2012PhRvD..85b7301L,2012PhRvL.108w1101H} had made stringent constraints on GRB physics, e.g., large Lorentz factor, large radius of emission region and/or small baryonic loading factor \citep{ICGRB2017,2012PhRvL.108w1101H,2012ApJ...752...29H}; Thus, the non-detection of PeV neutrinos from GRBs by IceCube may not conclude that GRBs are not UHECR sources, since it could be that the PIs and hence neutrino production may be inefficient. Indeed this is an advantage for GRBs to be UHECR sources since the UHECRs do not lose energy significantly in sources.

While PeV neutrino production had been constrained by IceCube, there is no crucial observational constraint in the EeV energy scale. 
The next generation neutrino telescopes with large array using radio detection techniques, such as IceCube-Gen2 \citep{ICgen2} and GRAND \citep{GRAND_whitepaper}, are being developed, which are aiming for higher sensitivities for detecting EeV neutrinos. 
In theory side, GRBs are expected to produce neutrino with energy $\sim {\rm EeV}$ by UHECRs in the reverse shock region \citep{Waxman_EeVafterglow,2001ApJ...551..249D}, in long-term afterglows \citep[e.g.,][]{2002A&A...396..303L}, or through $\pi^0$ production and decay in the prompt emission phase \citep{2007arXiv0711.4969L}. The detection of EeV neutrinos from GRBs will give direct and crucial evidence of UHECR production in GRBs. On the other hand,
for neutrino production at high energies, say, far beyond PeV, the parent pions and muons may suffer strong cooling before decay, namely, the damping effect in the GRB prompt emission region. The pion damping will significantly suppress the high-energy neutrino flux \citep{1998PhRvD..59b3002W}, whereas the muon damping modifies the neutrino flavor ratio in the source \citep{KashtiWaxman_flavor}. The observation of UHE, $\sim$EeV, neutrinos is important probe of GRB physics since the damping depends on the physical condition of the GRB emission region.

In this work, we study the prompt high energy neutrino emission from the GRB jets, focusing on the high energy part of the spectrum and the flavor fraction, which may suffer from strong damping effect. Different from the usual approach that somehow consider neutrino production within a dynamical time of the expanding jet (namely, "instantaneous approximation" hereafter; see Appendix \ref{sec:instan approx}), we consider the evolution of the particle distribution and neutrino production rate, when the jet is expanding to radii much larger than the energy dissipation radii. This is important for the production of high-energy neutrinos as the relativistic expansion of jet may have impact on the damping effect, e.g., if pions and muons do not cool significantly within the limited time due to the jet expansion. We set up a model to numerically calculate the time-integrated neutrino flux from expanding GRB jets, with the damping effect and neutrino production at large radii taken into account. 

The paper is organized as follows: The method for calculating PIs and the time integration of neutrino production will be presented in Section \ref{sec:model}; Numerical results of the neutrino spectrum and flavor fraction are shown in Section \ref{sec:result}; Then conclusions and discussions are given in Section \ref{sec:discussion}.

\section{Model} \label{sec:model}

We build a model for PIs in GRBs and calculate the neutrino production from charged pion decay. A simplified model is adopted from \cite{HummerSimB} (H10) for PIs, which helps to efficiently calculate the spectrum of secondary particles, such as pions and muons, crucial for exploring the damping effect. In the model, the continuity equations for secondary pions, muons, and hence neutrinos are solved, with the processes of production, cooling, and decay taken into account.

\subsection{GRB jet and radiation}\label{sec:2.1}

Consider that a relativistic jet with Lorentz factor $\Gamma$ is launched from the central engine of GRB. As the jet expands, the distance of jet from the central engine increases with time as $R\simeq \Gamma c t$, where $t$ is the time measured in the rest frame of jet. The thickness of the jet in the source frame is roughly $\Delta\simeq cT^{\rm ob}$, where $T^{\rm ob}$ is the observed duration of the GRB\footnote{Hereafter, the up-script "ob" denotes the quantity measured in the observer frame.}. When the jet expands to $R=R_0\gg\Delta$, corresponding to a time $t_{0} = R_{0}/ \Gamma c$, energy dissipation of the jet occurs, which produces cosmic rays and prompt gamma-ray emission. Afterward, a ``shell" of gamma-ray photons propagate with the jet, and the photons continuously interact with cosmic rays within a jet crossing time of $\Gamma\Delta/c$ in the rest frame of jet, after which photons escape and go ahead of the jet. In the jet crossing time, the jet propagate to $R\simeq R_0+\Delta R\simeq\Delta R\simeq c\Gamma(\Gamma\Delta/c)\simeq\Gamma^2cT^{\rm ob}$, beyond which the PI interaction and neutrino production basically end.\footnote{Usually, only neutrino production within a dynamical time $t$ of jet, i.e., around $R_0$ is considered. However, in this work, we consider neutrino production in much longer time $\Gamma\Delta/c$, i.e., up to radius $\Gamma^2cT^{\rm ob}$.}

The dissipation radius $R_0$ depends on theoretical models of jets. In the kinetic energy dominated jet model, it is expected that collisions between different parts of the jet due to different velocities result in internal shocks \citep{1993ApJ...418L...5P,1994ApJ...430L..93R} at radius $R_0\simeq\Gamma^2c\delta t$, where $\delta t$ is the observed variability time reflecting the variability of the central engine. Typically, $R_0\sim10^{14}$cm for $\Gamma\sim400$ and $\delta t\sim 10$ms. On the other hand, in the magnetically dominated jet model \citep{1997ApJ...482L..29M,2003astro.ph.12347L,2011ApJ...726...90Z}, the energy dissipation due to, e.g., magnetic reconnection, is expected to take place at a larger radius, $R_0\ga10^{16}$cm.

The observed spectrum of the prompt gamma-ray photons usually follows the Band function \citep{1993ApJ...413..281B}, i.e., a low-energy power-law $\epsilon^{-\alpha}$ transits smoothly to the other power law at high energies, $\epsilon^{-\beta}$, around the break energy $\epsilon_b$. Here $\epsilon$ and $\epsilon_b$ are measured in the rest frame of jet, where the photons can be approximated to be homogeneous and isotropic. Due to the expansion of jet, the photon density drop as $\propto R^{-2}$. In the rest frame of jet, the number density of the prompt emission photons per unit photon energy at $R$ can be described as
\begin{equation}\label{eq:n_gamma}
    n_{\gamma}(\epsilon,R) = 
    \begin{cases} 
        n_0 \left(\frac{R}{R_{0}}\right)^{-2} \left(\frac{\epsilon}{\epsilon_b}\right)^{-\alpha} &  \epsilon_{\min} \leq \epsilon \leq \epsilon_b \\
        n_0 \left(\frac{R}{R_{0}}\right)^{-2} \left(\frac{\epsilon}{\epsilon_b}\right)^{-\beta} &  \epsilon_b \leq \epsilon \leq \epsilon_{\max}
        \\
        0 & \rm else
    \end{cases},
\end{equation}
where $n_0 = n_\gamma(\epsilon_b,R_0)\simeq L_{\gamma} / (4 \pi R_{0}^{2} c \Gamma^2 \epsilon_{b}^{2})$, with $L_\gamma$ the GRB luminosity, and $\epsilon_{\min}$ and $\epsilon_{\max}$ are the low- and high-energy cutoffs, expected e.g., due to synchrotron self-absorption and photon-photon pair production, respectively. 

We will take the typical values,  $\alpha=1$, $\beta=2$, $\epsilon_b^{\rm ob}=\Gamma\epsilon_b=1$\,MeV, $L_\gamma=10^{52}$erg\,s$^{-1}$, and $T^{\rm ob}=10$\,s from observations, and assume $\epsilon_{\min}^{\rm ob}\sim0.1$\,keV \citep{residual_collision}, and $\epsilon_{\max}^{\rm ob}\sim10$\,GeV \citep[e.g.,][]{2007arXiv0711.4969L}, expected in GRB models.

Magnetic field plays an important role in GRB energy dissipation region, since the prompt emission is believed to be dominated by synchrotron radiation by accelerated electrons. The magnetic field is expected to envolve as the jet expands. We assume after $R_0$, the magnetic strength $B$ in the rest frame of jet decays with increasing $R$ as,
\begin{equation}\label{eq:B_field}
    B(R) = B_{0} \left(\frac{R}{R_{0}}\right)^{-b}.
\end{equation}
Defining $\xi_B$ the magnetic to gamma-ray energy ratio at $R_0$, $B_{0}$ can be given by $B_{0}\simeq (2\xi_B L_{\gamma}/R_{0}^{2} \Gamma^{2} c)^{1/2}$. The power-law index $b$ describes the evolution of magnetic field strength, i.e., $b=1$ for magnetic flux conservation, and $b=3/2$ for magnetic energy conservation.

Cosmic rays are only produced at the energy dissipation radius $R_0$ by particle acceleration processes. Afterward the cosmic rays are expected to be confined and propagate together with the jet. Assume that proton is the dominated composition of the produced cosmic rays, and as usual, assume the accelerated protons follow a flat power law with a sharp cutoff. Thus, in the rest frame of jet, the number of accelerated protons per unit proton energy $E_p$ at $R_0$ can be written as
\begin{equation}\label{eq:Np_0}
    N_{p}(E_{p},R_{0})= 
    \begin{cases}
        N_{0} E_{p}^{-2} & E_{\min} < E_{p} < E_{\max},
        \\
        0 & \rm else.
    \end{cases}
\end{equation}
We take $E_{\max}^{\rm ob} = \Gamma E_{\max} = 10^{21}$eV as the maximum energy, and the minimal energy should be larger than proton rest mass, $E_{\min} \gtrsim 1 {\rm GeV}$. Here the normalization $N_{0}$ should satisfy $E_{\gamma} f_{p} =\Gamma \int_{E_{\min}}^{E_{\max}} E_{p} N_{p}(E_p,R_0) \, {\rm d} E_{p}$, where $E_\gamma\simeq L_\gamma T^{\rm ob}$ is the gamma-ray energy of the GRB and $f_{p}$ is the baryonic loading factor, e.g., the ratio between the energies of accelerated protons and gamma-ray emission. The spectrum of protons will evolve due to energy loss, as discussed in section \ref{sec:2.2}.

\subsection{Photopion interactions and nucleon spectral evolution}\label{sec:2.2}

In the rest frame of jet, the PIs between the accelerated protons and the photons of prompt emission leads to pion production and hence neutrino production from pion decay.  There are several channels of pion production in PIs. One of the main channels is the $\Delta$ resonance, $p + \gamma \rightarrow \Delta^{+} \rightarrow p^{\prime} + \pi$, where the produced neutrinos carry an energy  of $E_{\nu} \sim E_{p}/20$ each, and there is an approximate relation between the proton energy and the target photon energy, $E_{p}^{\rm ob} \epsilon^{\rm ob} \simeq 0.2 \Gamma^2 \, {\rm GeV}^{2}$. However, for neutrino productions at high energies, the contribution from multi-pion channel becomes more important. In order for careful calculation of high energy neutrino productions, we should consider all production channels, especially the multi-pion chancel. Here we follow a simplified model, namely model \textit{Sim-B} from H10, for the PI calculation, containing resonant, direct and multi-pion channels.

Once provided the proton and photon spectra, $N_{p}(E_p,R)$ and $n_\gamma(\epsilon,R)$, respectively, we can calculate the production rate of charged pions per unit energy interval at pion energy $E_\pi$ and radius $R$, $Q_{\pi}(E_{\pi},R)$, via Equation (\ref{eq: Qpi}) in Appendix \ref{sec:pion production} for $R\leq\Gamma^2cT^{\rm ob}$. The calculation of the production rate of charged pions sums up all $\pi^{+}$'s and $\pi^{-}$'s, because we only concern the total production including both neutrinos and anti-neutrinos. Moreover, in the PI process protons and neutrons may be converted to each other, but we treat protons and neutrons the same in calculating the pion production.

Note, we take $Q_{\pi}(E_{\pi},R) = 0$ for $R\ga R_{\rm over} = \Gamma^{2} c T^{\rm ob}$, because the number density of prompt photons in the jet drops significantly.

We should treat protons and neutrons together. The evolution of nucleon, including protons plus neutrons, spectrum, $N_p[E_p,R(t)]$, should be solved out by the continuity equation for nucleons, with the initial proton spectrum given in Equation (\ref{eq:Np_0}), 
\begin{equation}\label{eq:Np}
    \frac{\partial}{\partial t} N_{p} + \frac{\partial}{\partial E_{p}} \left(\dot{E}_{p} N_{p}\right) = 0.
\end{equation}
where $\dot{E}_p$ is the energy loss rate of protons. We consider particle cooling due to PIs of protons and neutrons, adiabatic expansion of plasma and synchrotron radiation; thus, $\dot{E}_{p}= - E_{p} t_{p, \rm cool}^{-1}$, with the proton cooling time (inversed) being
\begin{equation}\label{eq:proton-cooling}
     t_{p, \rm cool}^{-1}(E_p,R) = t_{p\gamma}^{-1}(E_p,R) + t_{p,\rm ad}^{-1}(E_p,R) + t_{p,\rm syn}^{-1}(E_p,R). 
\end{equation}
The first term in the right hand side accounts for PIs. As we treat protons and neutrons the same, the PI is simplified to be a cooling process. The cooling rate $t_{p\gamma}^{-1}(E_{p},R)$ is given in Appendix \ref{sec:pion production}. 

Next, consider adiabatic cooling. For charged particles confined in the expanding plasma, the adiabatic cooling timescale should be about the dynamical timescale, $t_{\rm ad}\simeq t_{\rm dyn} = R / \Gamma c$. However, neutrons are produced in the PIs and, without charged, are not confined in the plasma. In the case of efficient PIs with $t_{0} \gg t_{p\gamma}(R_{0})$, there will be about half of secondary nucleons that are neutrons; on the contrary, when $t_{0} \ll t_{p\gamma}(R_{0})$, nearly all the nucleons are protons. Therefore, we modify the adiabatic cooling rate by a factor of  $1-f_{\pi}(E_p)/2$, where $f_{\pi}(E_{p}) = \min\{t_{0}/t_{p\gamma}(E_{p},R_{0}), 1\}$, i.e.,
\begin{equation}\label{eq:proton_ad_time}
    t_{p,\rm ad}^{-1}(E_p,R)=\left[1-\frac{f_{\pi}(E_p)}2\right] t_{\rm dyn}^{-1}(R).
\end{equation}
Finally, the last term is for synchrotron cooling of protons and neutrinos, which indeed has negligible effect under the physical condition concerned (see section 3).

\subsection{Time evolving pion and muon spectra and neutrino production} \label{sec:2.3}

The charged pions, $\pi^{\pm}$'s, produce muons and electron neutrinos in their chain decays: $\pi \rightarrow \mu + \nu_{\mu}$; and $\mu \rightarrow e + \nu_{\mu} + \nu_{e}$. Consider the production, decay and cooling processes of pions and muons in the rest frame of the GRB jet. The evolution of the pion and muon spectra,  $N_{\pi}(E_{\pi},R)$ and $N_{\mu}(E_{\mu},R)$, respectively, should be solved out by their continuity equations,
\begin{equation}\label{eq:pi_mu_CE}
    \frac{\partial}{\partial t} N_{x} + \frac{\partial }{\partial E_{x}} \left(\dot{E}_{x} N_{x}\right)
    = Q_{x} - \frac{N_{x}}{\tau_{x}},
\end{equation}
where $x=\pi$ or $\mu$, $Q_{x}$ is the production rate of pions or muons, $\tau_{x}$ is the decay timescale of pions or muons, and $\dot{E}_{x}$ is the energy-loss rate. Synchrotron and adiabatic cooling is the dominant processes for the energy loss of pions and muons, thus we have $\dot{E}_{x} = - E_x/t_{x, \rm cool}$, where the cooling time is given by
\begin{equation}
    t_{x, \rm cool}^{-1}(E_x,R)=t_{x,\rm syn}^{-1}(E_x,R)+t_{\rm dyn}^{-1}(R),
\end{equation}
with the synchrotron cooling time being $t_{x,\rm syn}^{-1} = c \sigma_{{\rm T},x} E_{x} B^{2}/6 \pi m_{x}^{2} c^{4}$, where the Thomson cross section for pions or muons is $\sigma_{{\rm T},x} = (8/3)\pi (e^2/m_{x}c^2)^2$, with $m_x$ the rest mass. We should ignore inverse Compton (IC) scatterings for energy loss of particles here, because for muons and pions with energy $E_x^{\rm ob}= \Gamma E_{x} \gtrsim 100 {\rm PeV}$,
the factor $E_{x}\epsilon_{b}/(m_{x} c^2)^2 \gtrsim 10$, indicating strong Klein-Nishina suppression so that IC cooling is much weaker than synchrotron cooling. As for the decay of pions and muons, the decay times in the rest frame of jet are $\tau_{\pi} = E_{\pi}\tau_{\pi,0} / m_{\pi}c^{2}$ and $\tau_{\mu} = E_{\mu}\tau_{\mu,0} / m_{\mu}c^{2}$, where $\tau_{\pi,0} = 2.6 \times 10^{-8} {\rm s}$ and $\tau_{\mu,0} = 2.2 \times 10^{-6} {\rm s}$ are the life times of corresponding particles in their rest frame.

With the evolution of pion production rate $Q_\pi$ (given in Section \ref{sec:2.2} and Appendix \ref{sec:pion production}), one can derive pion spectrum $N_\pi$ from Equation (\ref{eq:pi_mu_CE}) (with $x=\pi$). 

Next, consider muon production rate $Q_\mu$, as well as the production rates of neutrinos $Q_{\nu_\mu}$ and $Q_{\nu_e}$. They are all secondaries from decay of some parent particles. In a decay chain, $a\rightarrow b$, where the secondary particle $b$ is generated by the decay of ultra-relativistic parent particle $a$, the production rate of $b$, $Q_b=Q_{a\rightarrow b}$, can be calculated as
\begin{equation}\label{eq:Qmu,Qnu}
    Q_{a \rightarrow b}(E_{b},R) = \int {\rm d} E_{a} \frac{N_{a}(E_{a},R)}{\tau_{a}(E_{a})} p_{a \rightarrow b}(E_{b};E_{a}),
\end{equation}
where $\tau_{a}$ is the decay time of parent particles (in the rest frame of jet), and $p_{a \rightarrow b}(E_{b};E_{a})$ is the probability distribution function of secondary particle $b$. 
The forms of $p_{\pi \rightarrow \mu}$, $p_{\pi \rightarrow \nu_\mu}$, $p_{\mu \rightarrow \nu_\mu}$, and $p_{\mu \rightarrow \nu_e}$ are shown in Appendix \ref{sec:decay secondary distribution}. Note, here we concern only the total flux of neutrinos and anti-neutrinos and will not treat them separately.  
With the derived pion spectrum, $Q_\mu$ can be obtained (Equation  \ref{eq:Qmu,Qnu}), and inserted into Equation (\ref{eq:pi_mu_CE}) ($x=\mu$) to derive the evolving muon spectrum. Hence, the production rates of electron and muon neutrinos, $Q_{\nu_{e}}(E_{\nu},R)=Q_{\mu \rightarrow \nu_e}(E_\nu, R)$ and $Q_{\nu_{\mu}}(E_{\nu},R)=Q_{\pi \rightarrow \nu_{\mu}}(E_\nu, R) + Q_{\mu \rightarrow \nu_{\mu}}(E_\nu, R)$, respectively, can be calculated by Equation (\ref{eq:Qmu,Qnu}) using already derived pion and muon spectra. 

The neutrino spectra at radius $R$ should be integration from initial time $t_0$ up to time $t(R)=R/\Gamma c$. The muon and electron neutrino spectra (neutrino number per unit neutrino energy in the rest frame of jet) at $R$ are, respectively, 
\begin{equation}
    N_{\nu_{\mu}}(E_{\nu},R) = \int_{t_0}^{t(R)} {\rm d}t_i  Q_{\nu_{\mu}}(E_\nu, R_i),
\end{equation}
and
\begin{equation}
    N_{\nu_{e}}(E_{\nu},R) =  \int_{t_0}^{t(R)} {\rm d}t_i\,  Q_{\nu_{e}}(E_\nu, R_i),
\end{equation}
with $R_i = \Gamma c t_i$. To calculate the total neutrino number spectrum, $N_\nu(E_\nu)$, the upper bound of the integrals in the above two equations should be the time (or radius) where all the produced muons decay. The observed time-integrated neutrino flux as function of neutrino energy, i.e., fluence spectrum, can be obtained, 
\begin{equation}
    \Phi_{\nu}(E_{\nu}^{\rm ob}) = \frac{N_{\nu}(E_{\nu})}{\Gamma4 \pi d_L^{2}},
\end{equation}
where $E_\nu=E_\nu^{\rm ob}/\Gamma$, and $d_L$ is the luminosity distance of the GRB.

In Equation (\ref{eq:pi_mu_CE}), both radiative cooling and decay have been taken into account in the evolution of pion and muon distributions, so the damping effects have been considered for neutrino productions. However, different from the approach of instantaneous approximation (see Appendix \ref{sec:instan approx} for details) for pion and muon distributions, we concern their time-evolution during the jet expansion, and obtain the final neutrino spectrum and flavor fraction by summing up contributions of all time and radii.

\section{Result} \label{sec:result}

To solve the continuity equations, Equations (\ref{eq:Np}) and (\ref{eq:pi_mu_CE}), for the time-dependent spectra of muons and pions, we use the fully implicit difference scheme \citep{FIDS}, which can find stable and accurate solutions with less mesh points.
Hereafter we will take the following parameter values as fiducial ones, unless stated otherwise: $\Gamma=400$, $R_{0} = 10^{14}$cm, $b=1$, $\xi_B=1$, and $f_p=10$.

\begin{figure}[t]
    \centering
    \includegraphics[width=1\linewidth]{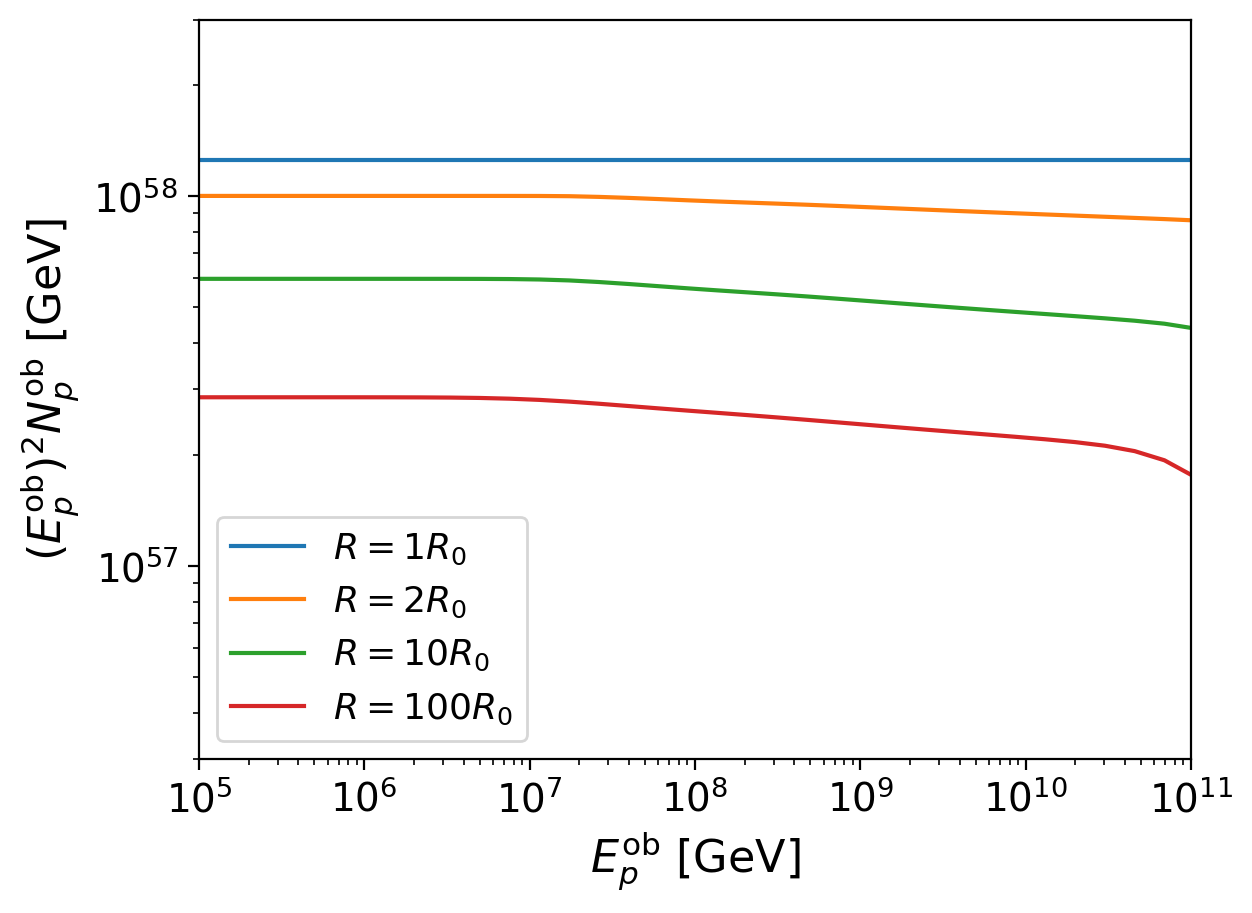}
    \includegraphics[width=1\linewidth]{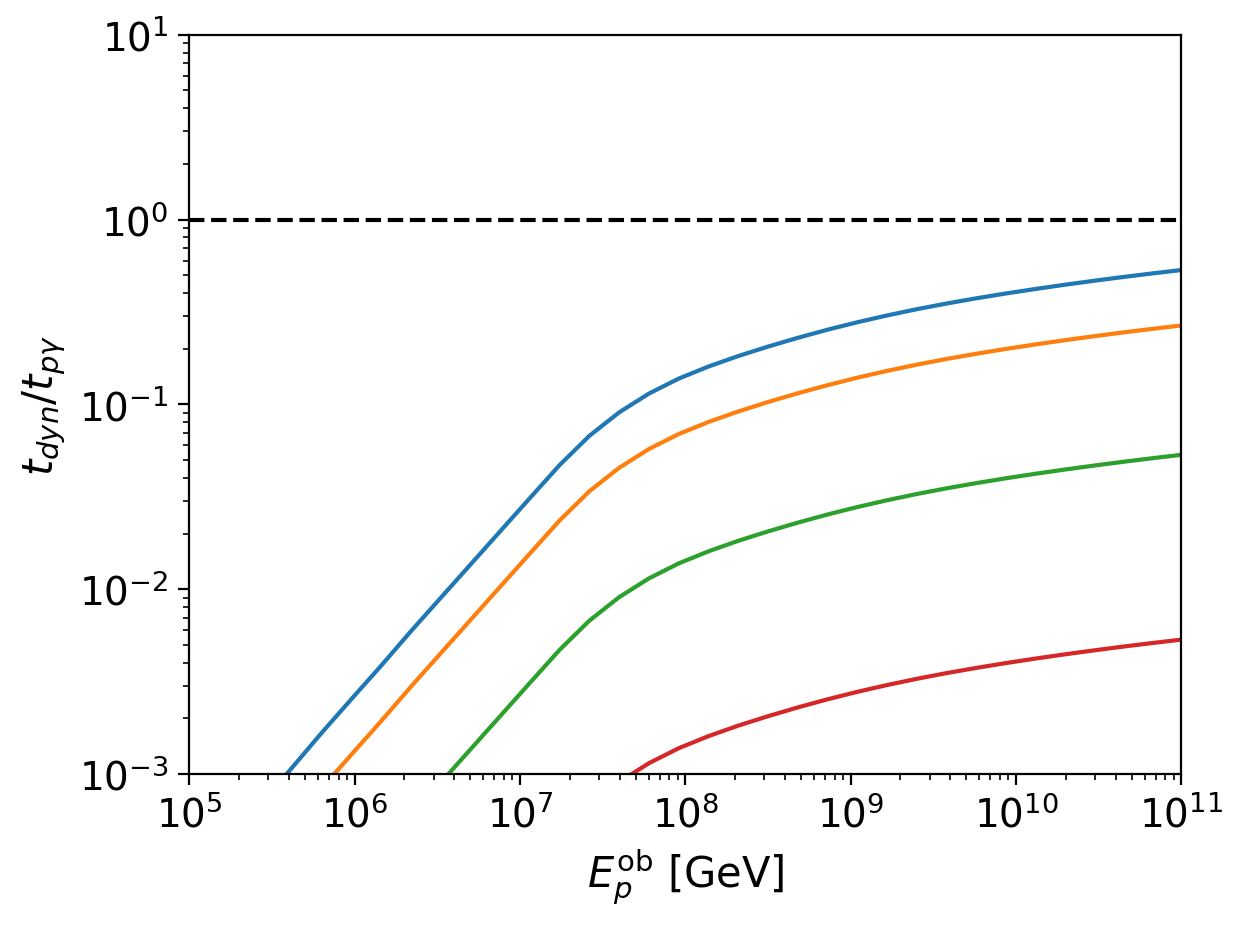}
    \caption{The nucleon spectrum (upper panel) and the ratio $t_{\rm dyn} / t_{p\gamma}$ as function of nucleon energy (lower panel), evolving with jet radius $R$. Fiducial parameter values are used (see text). }
    \label{fig:proton1}
\end{figure}
\begin{figure}[t]
    \centering
    \includegraphics[width=1\linewidth]{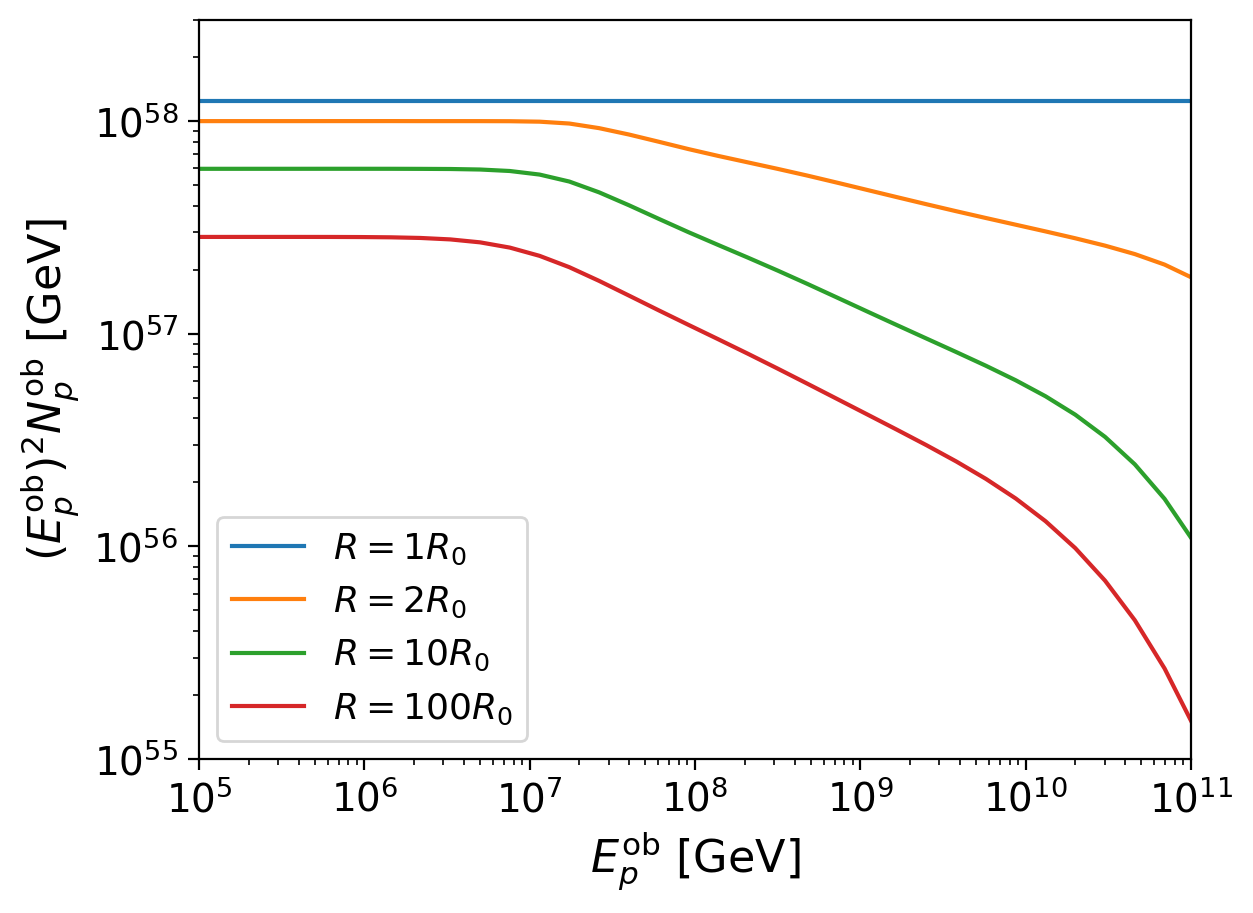}
    \includegraphics[width=1\linewidth]{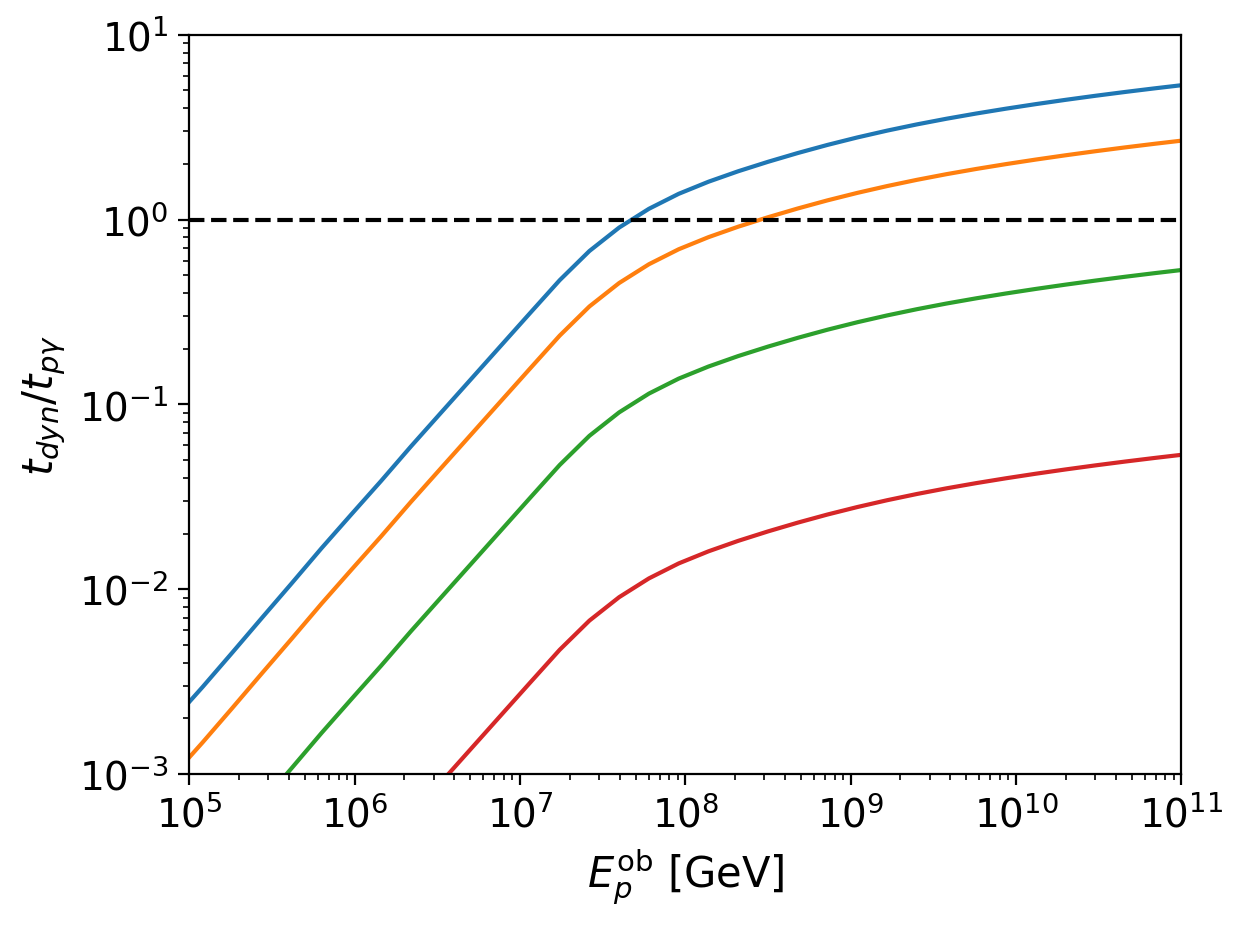}
    \caption{Same as Figure \ref{fig:proton1}, but with $R_{0} = 10^{13} {\rm cm}$.}
    \label{fig:proton2}
\end{figure}

\subsection{Time-dependent spectra of parent and secondary particles}\label{sec:evolution}

We first show the results about how the spectra of parent particles and secondary products evolve with time, or radius as the jet is expanding.
Figure \ref{fig:proton1} presents the evolution of the spectrum of nucleons, i.e., protons plus neutrons, during the expansion of the jet in the case of fiducial parameter values. The upper panel shows the spectra of nucleons in the observer frame, $(E_p^{\rm ob})^2N_{p}^{\rm ob}(E_{p}^{\rm ob},R) = \Gamma E_p^2N_{p}(E_{p},R)$, at different radii. As the spectrum of nucleons are determined by the cooling process of neucleons, which is in term determined by the PI and adiabatic cooling (Equation \ref{eq:proton-cooling}), we present in the lower panel of Figure \ref{fig:proton1} the ratio $t_{\rm dyn}(R)/t_{p\gamma}(E_{p},R)$ at different radii. This ratio defines an "optical depth of protons" due to PIs. 


With fiducial parameter values, the optical depth is initially  smaller than unity across the energy range concerned, i.e., "optically thin" case. As photon number density decreases as $n_\gamma\propto R^{-2}$, $t_{p\gamma}^{-1}\propto R^{-2}$, and hence the ratio $t_{\rm dyn}/t_{p\gamma}\propto R^{-1}$ is decreasing. So this case is always with $t_{\rm dyn}/t_{p\gamma}<1$ (lower panel of Figure \ref{fig:proton1}), implying that the cooling by PIs does not affect the nucleon spectrum significantly. Thus, the adiabatic cooling should dominate the nucleon spectral evolution, leading to an energy-independent decrease of nucleon flux. As seen in Figure \ref{fig:proton1}, at $R\la2R_0$ the spectrum does not change significantly, and only at $R\ga10R_0$ the spectrum moves significantly below, but with the spectral shape largely unchanged.

For comparison, we show in Figure \ref{fig:proton2} the spectra of nucleons in an "optically thick" case with much smaller initial radius, $R_{0} = 10^{13}$cm. We can see in the lower panel of Figure \ref{fig:proton2} that the timescale ratio is initially larger than unity for nucleons with energy $E_{p}^{\rm ob} \gtrsim 10^{7} {\rm GeV}$, i.e., the cooling through PIs is significant for the highest-energy nucleons at the beginning. At $R=2R_0$ the strong cooling results in significant spectral steepening at high energies since the cooling rate $t_{p\gamma}^{-1}$ increases with $E_p$. High energy nucleons continue cooling fast by PIs until the ratio decreases to $t_{\rm dyn}/t_{p\gamma}<1$ across all nucleon energies at $R\ga10R_0$. After that, the nucleon number decreases with the spectral profile keeping unchanged (upper panel of Figure \ref{fig:proton2}). The much steeper nucleon spectrum at large radii compared to the optically thin case will result in suppression of neutrino production (discussed in Section \ref{sec:nu spectrum}).

\begin{figure}[t]
    \centering
    \includegraphics[width=1\linewidth]{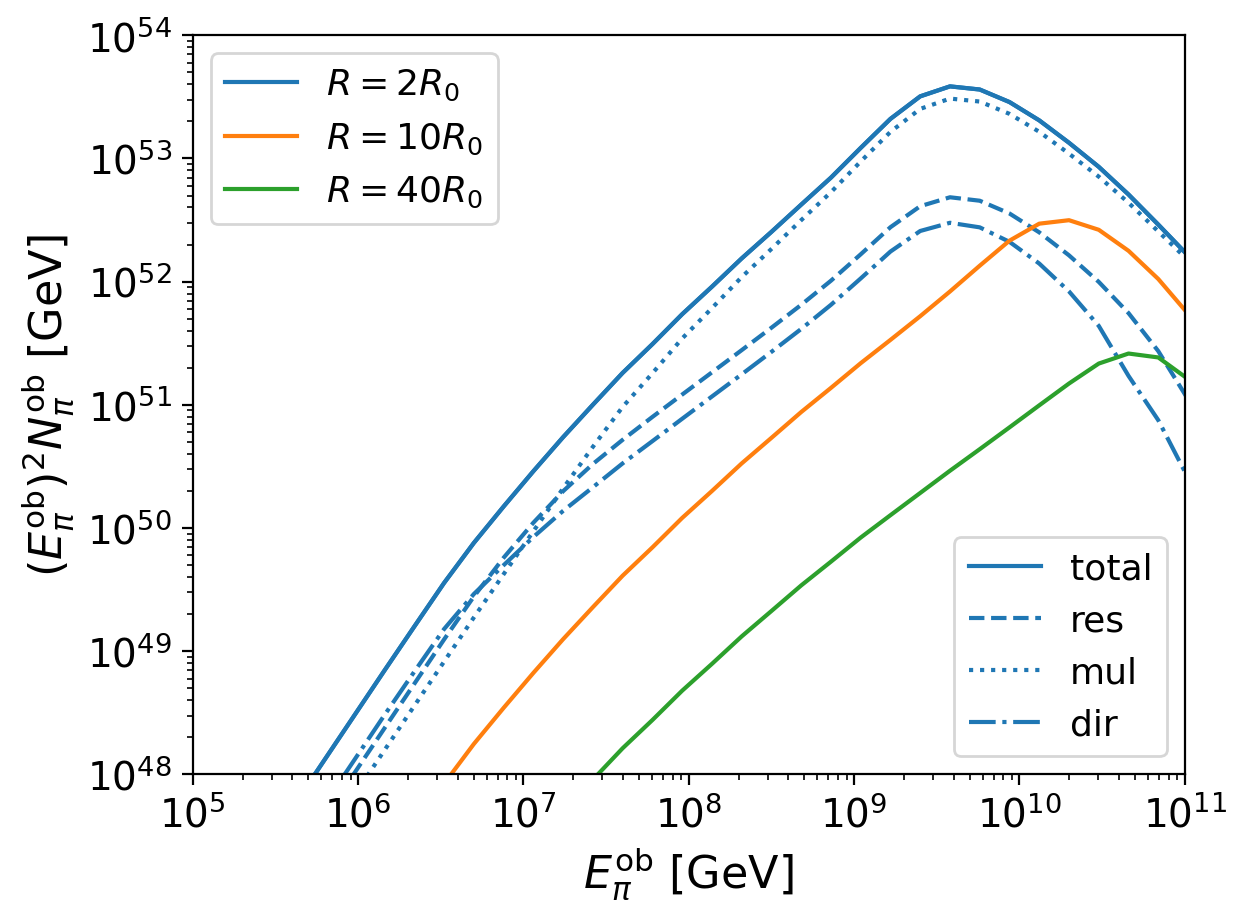}
    \includegraphics[width=1\linewidth]{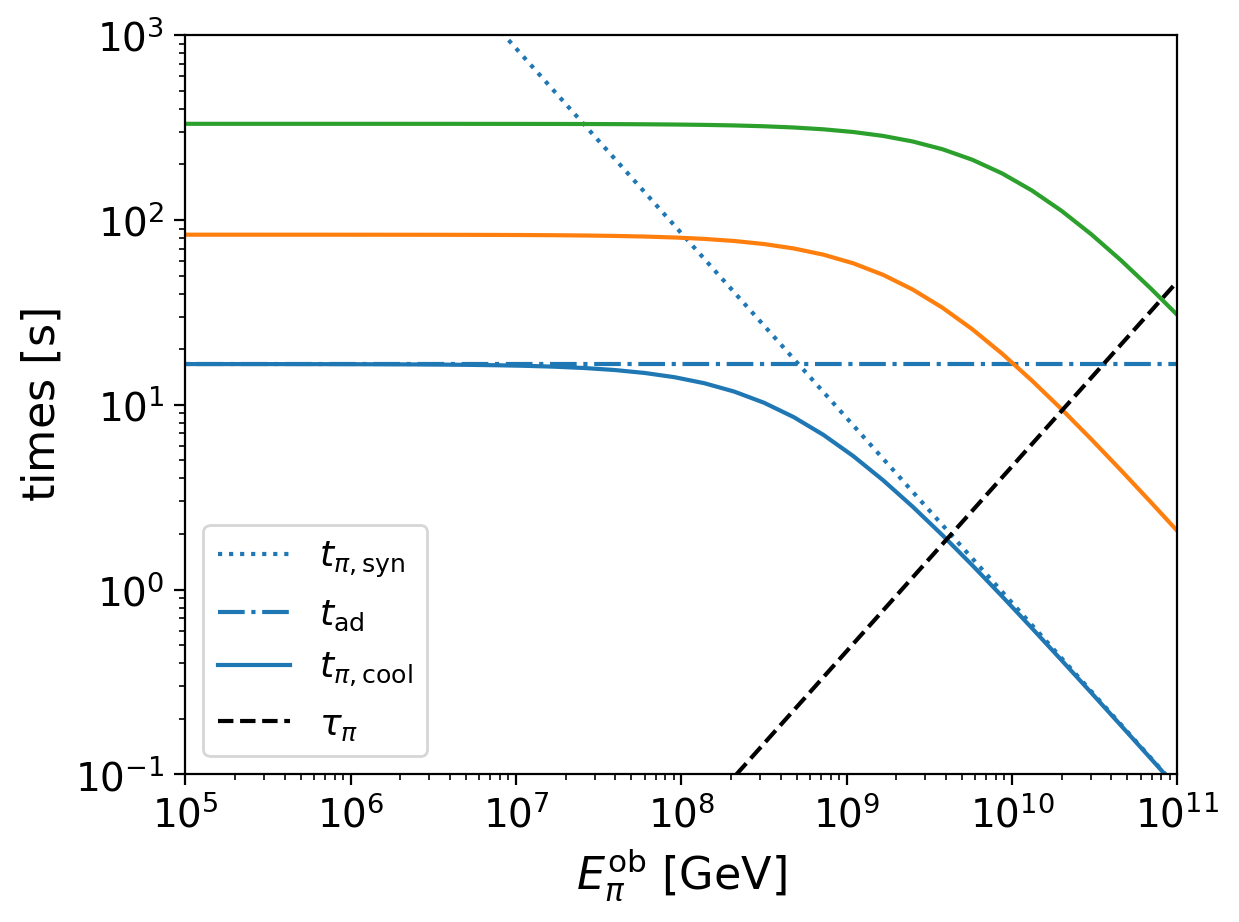}
    \caption{The charged pion spectrum (upper panel) and the cooling and decay time of charged pions as functions of pion energy (lower panel) evolving with jet radius $R$. For $R=2R_0$ the contributions of pion production by resonant, direct and multi-pion channels and the synchrotron and adiabatic cooling times of charged pions are shown for examples. }
    \label{fig:pion}
\end{figure}
\begin{figure}[t]
    \centering
    \includegraphics[width=1\linewidth]{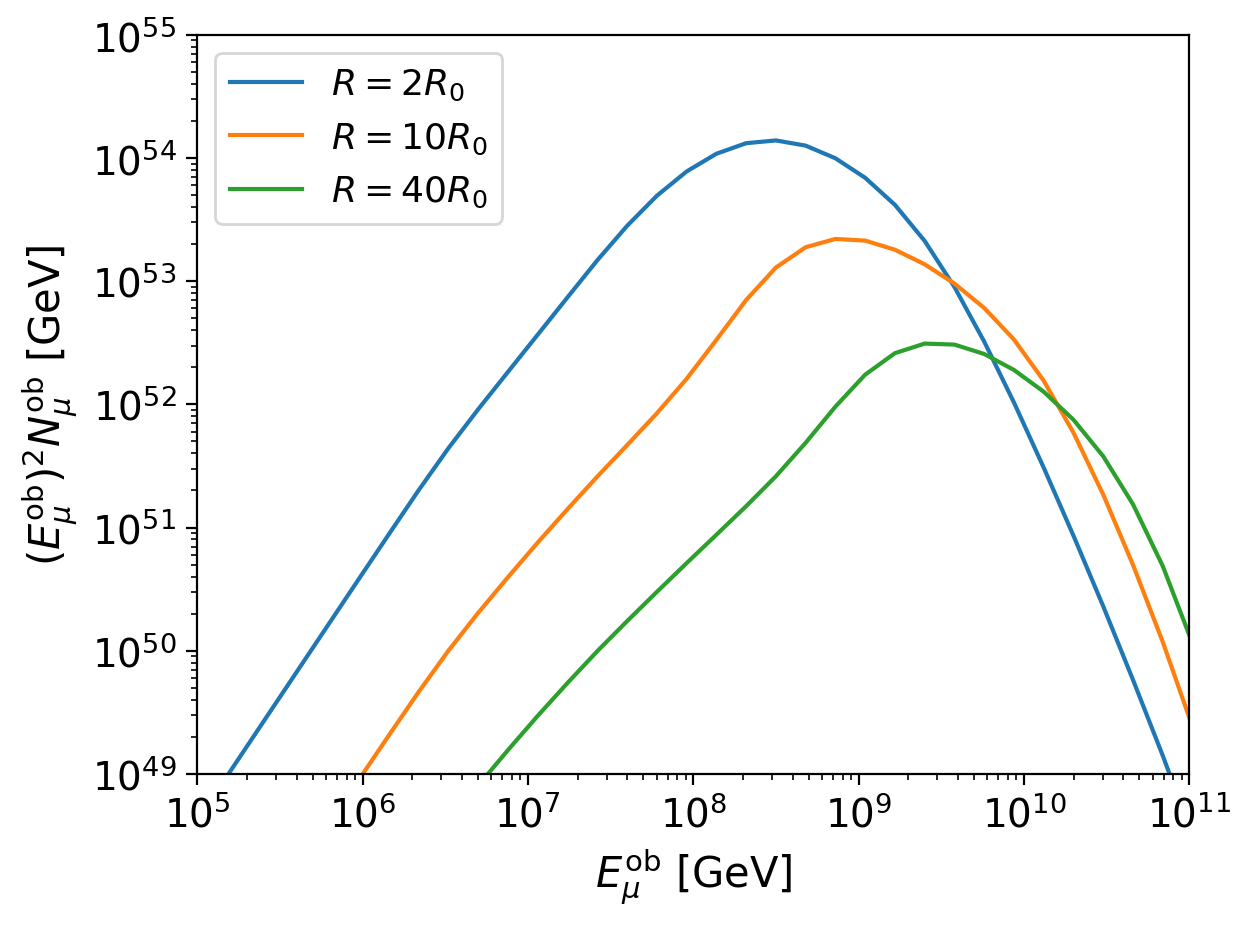}
    \includegraphics[width=1\linewidth]{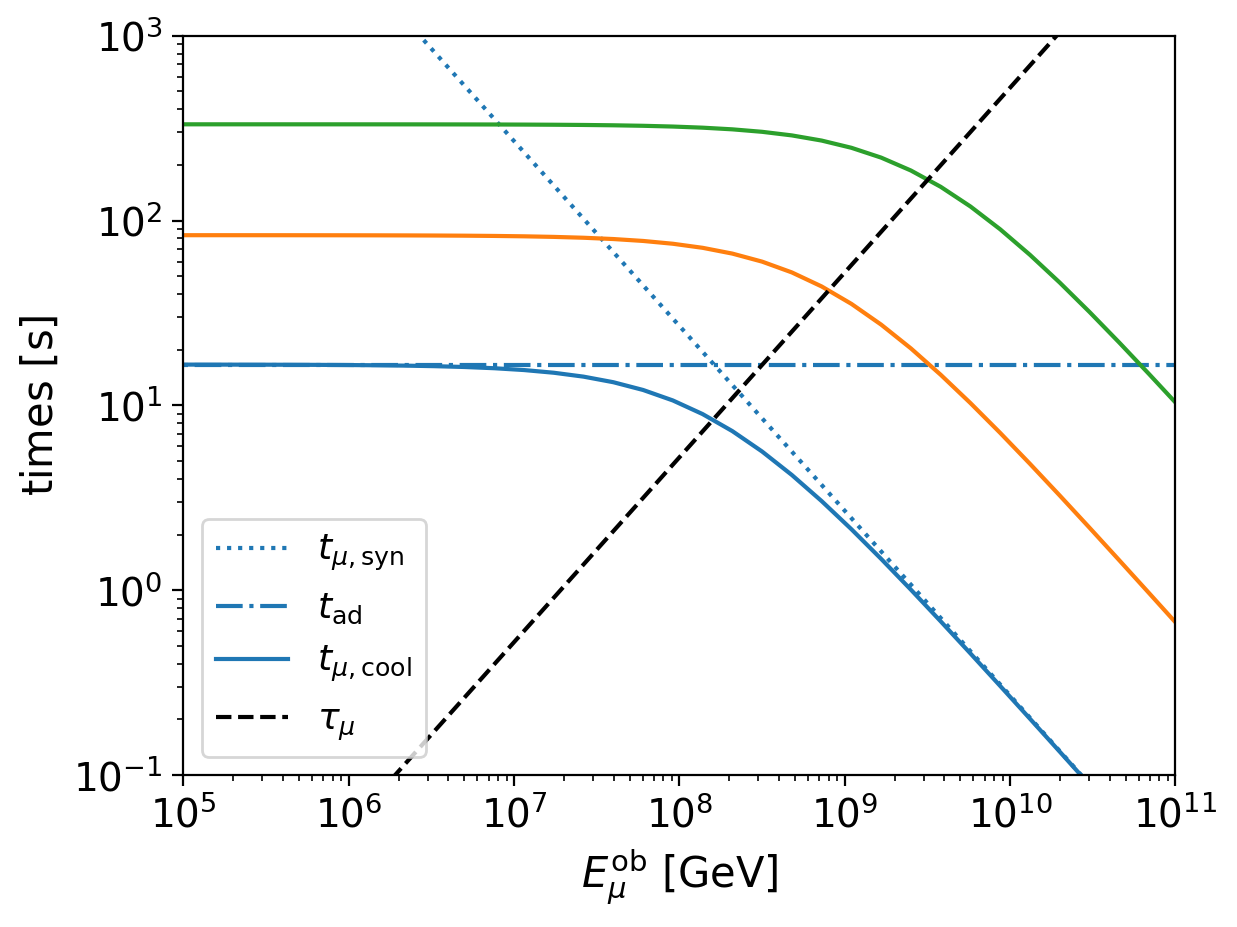}
    \caption{Same as Figure \ref{fig:pion}, but for muons.}
    \label{fig:muon}
\end{figure}

 
Next, we present in Figures \ref{fig:pion} and \ref{fig:muon} the evolution of the spectra of the produced pions and muons. For one pion spectrum in Figure \ref{fig:pion}, the contribution from different types of channels for pion production are shown. In the upper panel of Figure \ref{fig:pion}, a break around $E_{\pi,b}^{\rm ob}\sim10^{7}$GeV can be found in the pion spectrum from the resonant channel at $R=2R_0$. This break is caused by a corresponding spectral break $\epsilon_b$ in photon spectrum (Equation \ref{eq:n_gamma}), leading to  that the pion production rate from the resonant channel follows $E_{\pi}^2 (Q_{\pi})_{\rm res}\propto E_{\pi}^{\beta-1} \propto E_{\pi}$ for pions below the break energy and $E_{\pi}^2 (Q_{\pi})_{\rm res} \propto E_{\pi}^{\alpha-1} \propto E_{\pi}^{0}$ for above. The spectrum of produced pions is then $E_{\pi}^{2}(N_{\pi})_{\rm res} \sim E_{\pi}^{2}  (Q_{\pi})_{\rm res} \tau_{\pi} \propto E_{\pi}^{2}$ and $\propto E_{\pi}$, below and above the break energy, respectively. Above $E_{\pi,b}^{\rm ob}$ the pion contribution from the multi-pion channel, as can be seen, increases faster than that from the resonant and direct channels and becomes dominant at high energies. As a result, the total pion spectrum becomes more smooth around the spectral break at $E_{\pi,b}^{\rm ob}$.

As $t_{\rm dyn}/t_{p\gamma}\propto R^{-1}$, pion production is dominated by PIs around $R_0$ in either optically thin or optically thick cases. Afterward, the pion number decreases. In the upper panel of Figure \ref{fig:pion}, one sees that as the bulk pion number is decreasing with time/radius, there is a spectral peak in the pion spectrum evolving from $10^{9.5}$GeV to higher energies. These breaks are caused by pion damping, i.e., when the pion cooling timescale, $t_{\pi, \rm cool} = (t_{\pi, \rm syn}^{-1} + t_{\rm dyn}^{-1})^{-1}$, is smaller than the decay time of pions, $\tau_{\pi}$, pions significantly lose energy before decay. Denote $E_{\pi, \rm damp}^{\rm ob}(R)$ the pion damping energy, where the cooling and decay times are equal, $t_{\pi,{\rm cool}}= \tau_{\pi}$. As shown in the lower panel of Figure \ref{fig:pion}, due to the short decay time of pions, the damping energy is determined by comparing the synchrotron cooling time with the decay time. For the case of fiducial parameter values, the magnetic field decays as $B \propto R^{-1}$, thus, as shown in the lower panel, the pion damping energy at $R_0$ is $E_{\pi, \rm damp}^{\rm ob}(R_0) \sim 10^{9.5}$GeV \citep[which is consistent with other analytical estimate, e.g.][]{2012ApJ...746..164M}, and evolves as $E_{\pi,{\rm damp}}^{\rm ob} \propto R$. So the spectral break appears at higher energy for larger radius, and hence PIs at larger radii may significantly contribute to production of high energy neutrinos, see below.


Pion decay produce muons, whose spectral evolution is shown in Figure \ref{fig:muon}. Two spectral breaks show up in muon spectrum. The relatively low energy one, which is also the spectral peak, is due to muon damping, corresponding to the muon damping energy $E_{\mu,{\rm damp}}$, where the muon cooling time equal to muon decay time, $t_{\mu,\rm cool}=\tau_\mu$. 
As the muon decay time is much larger than that of pions, the muon damping energy is much lower than that of pions, $E_{\mu, \rm damp}^{\rm ob}(R_0) \sim 10^{8}$GeV (see the lower panel of Figure \ref{fig:muon}). One sees that at $E_{\mu,{\rm damp}}$  the muon synchrotron cooling time is comparable to the adiabatic cooling time, and hence adiabatic cooling cannot be ignored. However, for muons with larger energies, synchrotron cooling becomes dominant.

The other spectral break appears at relatively high energy in muon spectrum, beyond which the muon spectrum steepens sharply (upper panel of Figure \ref{fig:muon}). This break results from pion damping effect, $E_{\mu,c}\sim (3/4) E_{\pi, {\rm damp}}$, above which muon production is suppressed significantly and the flux decline with muon energy sharply due to pion and muon damping together. Note, in the upper panel of Figure \ref{fig:muon}, the muon flux at highest energies is larger at larger radius. This is because of the fast increase of pion damping energy $E_{\pi,\rm damp}\propto R$, and hence the break energy $E_{\mu, c}\propto R$, leading to more muon production at large radii. 

\begin{figure}[t]
    \centering
    \includegraphics[width=1\linewidth]{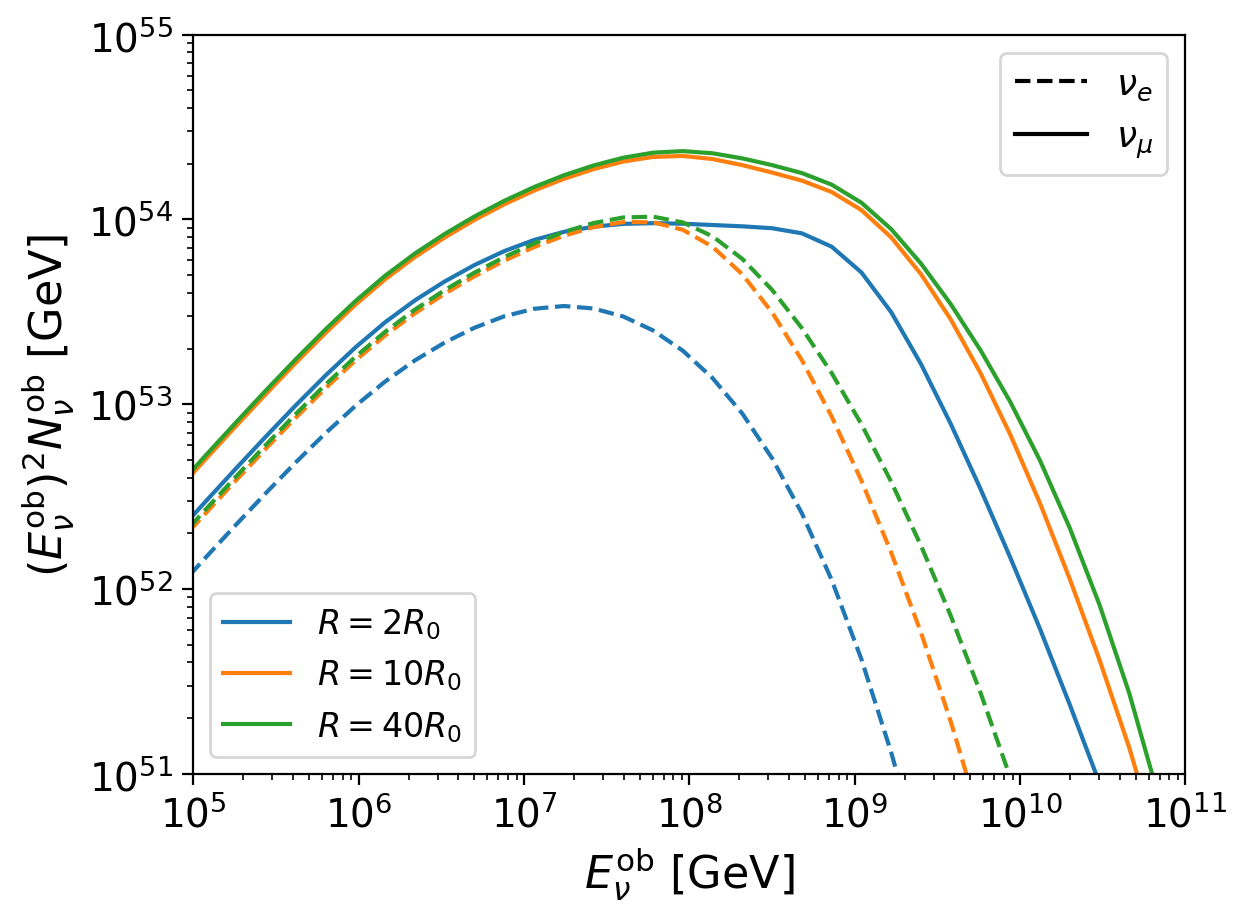}
    \caption{The neutrino spectrum evolving with jet radius $R$. 
    The dashed and solid lines represent electron and muon neutrinos, respectively.}
    \label{fig:neu_evolution}
\end{figure}

Finally, Figure \ref{fig:neu_evolution} shows the evolution of neutrino spectrum during the expansion of jet. We calculate the accumulative neutrino spectrum up to $R$, $N_{\nu}^{\rm ob}(E_{\nu}^{\rm ob},R)$, with the method described in section \ref{sec:2.3}. Since the interaction rate decreases rapidly with radius, $t_{p\gamma}^{-1} \propto R^{-2}$, the integration for neutrino number roughly gives $N_{\nu} \propto (1 - R_0/R)$, neglecting damping effect. So the neutrino production is almost completed at $R \sim 10 R_{0}$ for neutrinos with energy lower than the corresponding damping energies, which is the muon damping energy $E_{\mu,\rm damp}^{\rm ob}(R_0)/3 \sim 10^{7.5} {\rm GeV}$ for electron neutrinos and pion damping energy $E_{\pi,\rm damp}^{\rm ob}(R_0)/4 \sim 10^{9} {\rm GeV}$ for muon neutrinos. However, production at larger radii are still non-negligible for neutrinos with energy larger than the corresponding damping energy (see Figure \ref{fig:neu_evolution}).

\subsection{Neutrino fluence spectrum}\label{sec:nu spectrum}
\begin{figure}[t]
    \centering
    \includegraphics[width=1\linewidth]{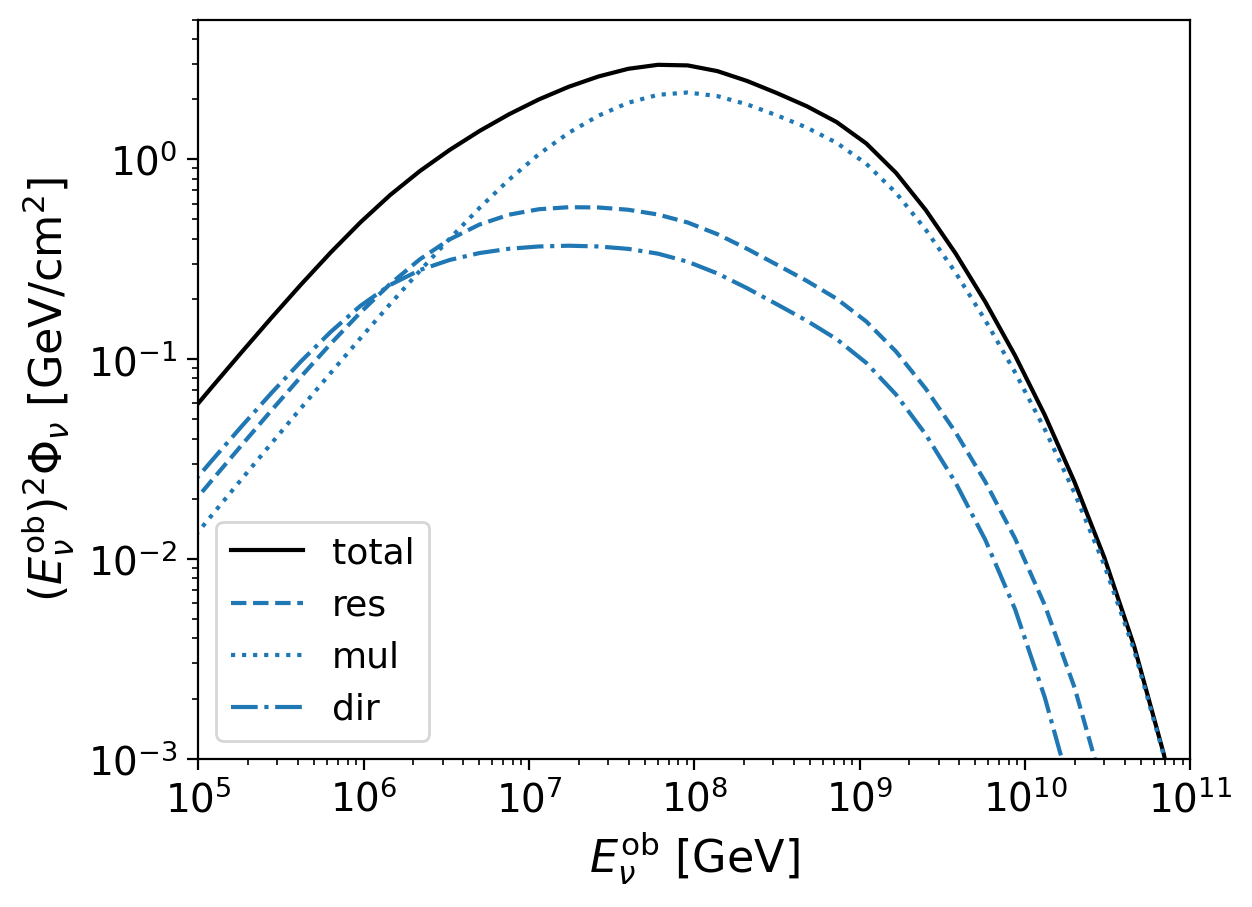}
    \caption{All-flavor neutrino fluence spectra: the total fluence (solid line) and the contributions from resonant (dashed line), multi-pion (dotted line) and direct (dash-dot line) channels.}
    \label{fig:channels}
\end{figure}

By integration up to a large radius, say, $R = 10^4 R_0$, we obtain the total neutrino fluence, assuming $d_L=100$\,Mpc hereafter. Figure \ref{fig:channels} shows the total fluence spectrum $(E^{\rm ob}_{\nu})^2\Phi_{\nu}$ of all-flavor neutrinos, as well as the contribution from different pion production channels. 
As well known, in the resonant channel, a spectral break corresponding to the gamma-ray spectral break, $\epsilon_{b}^{\rm ob}=1\epsilon_{b,\rm MeV}^{\rm ob}$MeV, should show up at neutrino energy $E_{\nu}^{\rm ob} \sim (1/20) \Gamma m_{p}c^{2} \epsilon_{\Delta} / 2 \epsilon_{b} \sim 10^{6} \Gamma_{400}^{2} (\epsilon_{b,\rm MeV}^{\rm ob})^{-1}$GeV ($\epsilon_{\Delta} \sim 0.3$ GeV for resonance), which can be seen in Figure \ref{fig:channels}. Below this break, the resonant and direct channels contribute most neutrinos and the fluence increases as $(E^{\rm ob}_{\nu})^2\Phi_{\nu} \propto E_{\nu}^{\rm ob}$, corresponding to the photon spectral slope of $n_{\gamma} \propto \epsilon^{-2}$ for $\epsilon > \epsilon_{b}$. From Figure \ref{fig:channels} one can find that the neutrinos with $E_{\nu}^{\rm ob} \gtrsim 10^{6.5} {\rm GeV}$ mainly result from the multi-pion channel. There are two more spectral breaks that appear in this energy range: one is $E_{\nu}^{\rm ob} \sim E_{\mu,{\rm damp}}^{\rm ob}(R_{0})/3 \sim 10^{7.5} {\rm GeV}$ (lower panel in Figure \ref{fig:muon}), corresponding to muon damping; the other is $E_{\nu}^{\rm ob} \sim E_{\pi,{\rm damp}}^{\rm ob}(R_{0})/4 \sim 10^{9} {\rm GeV}$ (lower panel in Figure \ref{fig:pion}), caused by pion damping. These features in the spectrum results from synchrotron cooling of charged particles, and the measurement of them could be important probes of the jet physics, e.g., the magnetic field.

\begin{figure}[t]
    \centering
    \includegraphics[width=1\linewidth]{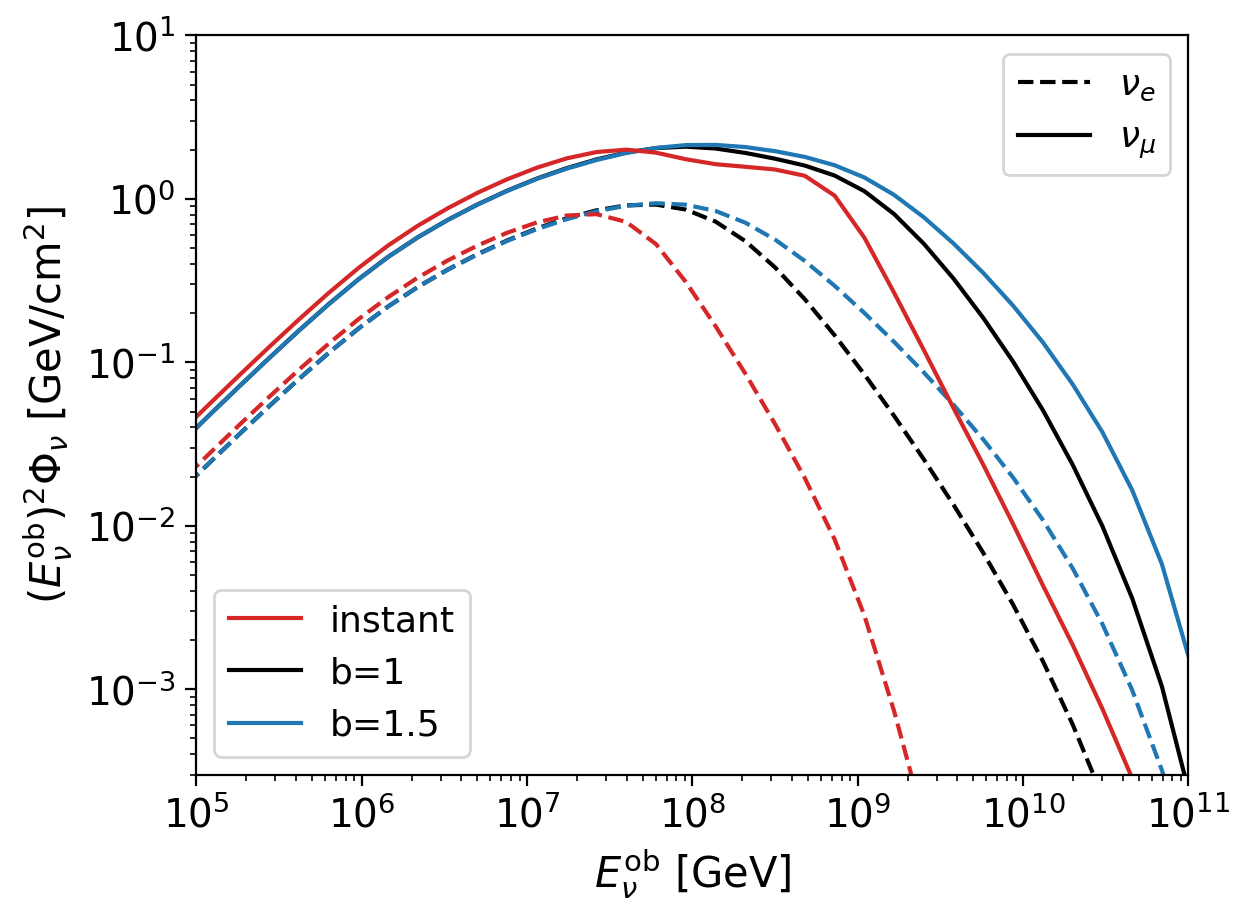}
    \caption{Neutrino fluence spectra with different magnetic field decay indices, $b=1$ (black line) and $1.5$ (blue line), in comparison with that of instantaneous approximation (red line). }
    \label{fig:magnetic_field_decay}
\end{figure}

In Figure \ref{fig:magnetic_field_decay} we present the impact of the magnetic field decay on neutrino spectra (before neutrino oscillation). We consider different decay index $b$ (Equation \ref{eq:B_field}), and calculate the neutrino spectrum under the instantaneous approximation for comparison. We have described in detail how the calculation is done in this approximation in  Appendix \ref{sec:instan approx}.  The approximation works well below $E_{\nu}^{\rm ob}\sim E_{\mu,{\rm damp}}^{\rm ob}(R_0)/3$, corresponding to muon damping, only with small difference due to the adiabatic cooling considered or not. The difference becomes significantly larger at higher energies. 
For electron neutrinos, the spectrum in instantaneous approximation decreases steeply, roughly as $(E_{\nu}^{\rm ob})^2 \Phi_{\nu} \propto t_{\mu,{\rm cool}}/\tau_{\mu} \propto E_{\nu}^{-2}$, in the energy range suffering muon damping, whereas the time-integrated spectra in this work is much shallower in the same energy range. For muon neutrinos, muon damping does not affect the spectrum significantly, but pion damping does significantly suppress the muon neutrino production. And the shape of muon neutrino spectra in the energy range suffering pion damping is similar to that of electron neutrinos suffering muon damping. 

This is the impact of neutrino contribution at large radii. As shown in Figure \ref{fig:neu_evolution}, the neutrino contribution at $R > 10 R_0$ becomes more important at higher energy. This is because that the damping energy increases as $\propto R^b$, then muons and pions produced at large enough radii is free from the damping effect, while they suffer strong damping effect around $R_0$. For larger $b$, the contribution at large radii is larger, leading to a shallower neutrino spectrum.

\begin{figure}[t]
    \centering
    \includegraphics[width=1\linewidth]{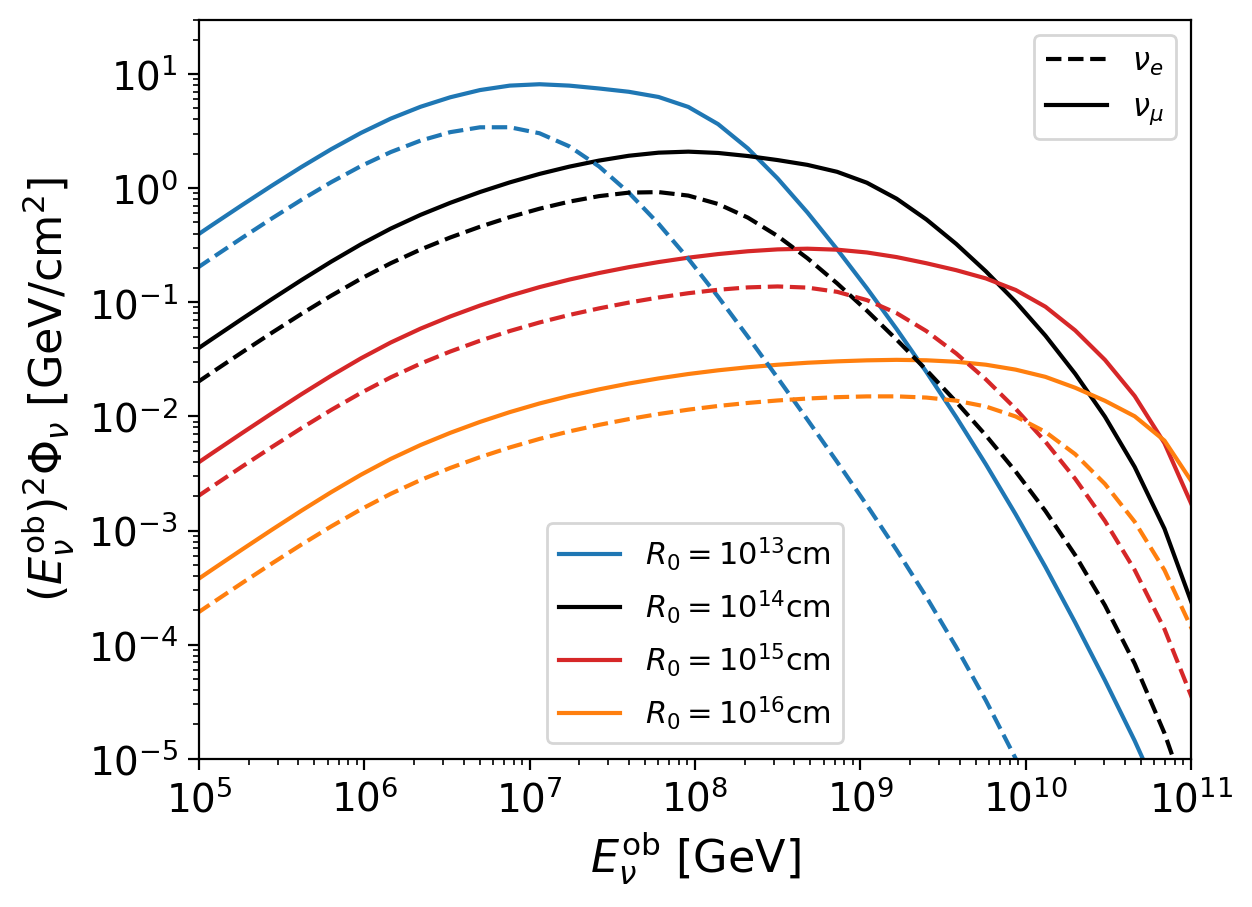}
    \caption{Neutrino fluence spectra with different initial radius $R_0$. }
    \label{fig:R0}
\end{figure}

\begin{figure}[t]
    \centering
    \includegraphics[width=1\linewidth]{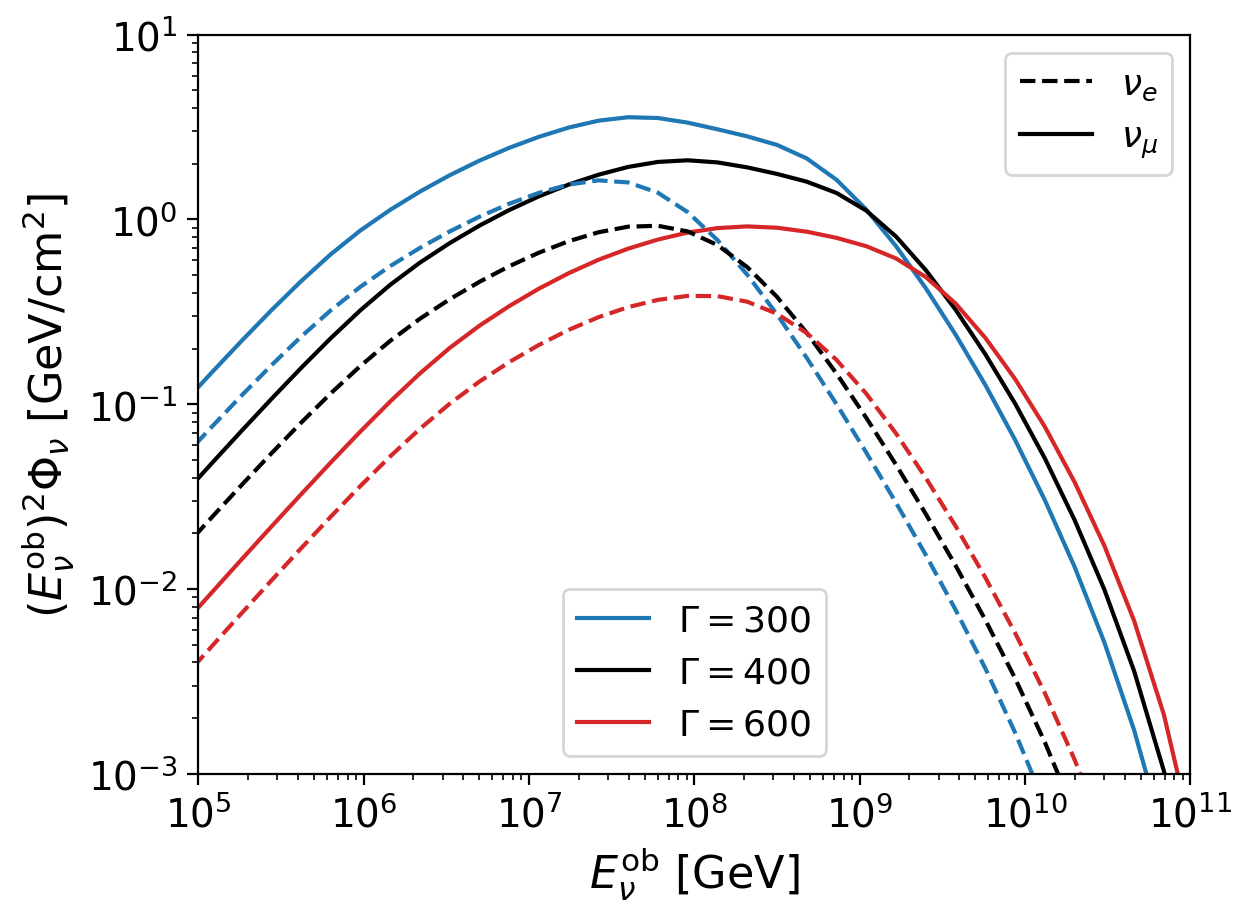}
    \caption{Neutrino fluence spectra with different jet Lorentz factor $\Gamma$. }
    \label{fig:Gamma}
\end{figure}

We further examine the effects of parameters $R_0$ and $\Gamma$ on the neutrino spectrum in Figures \ref{fig:R0} and \ref{fig:Gamma}. One can see in Figure \ref{fig:R0} as $R_0$ increases the neutrino fluence decreases and the spectrum extends to higher energies. Consider the cause below.  
As the photon number density decreases with $R_0$, $t_{p\gamma}^{-1} \propto R_{0}^{-2}$, and the low-energy neutrinos are mainly produced around $R_0$, with dynamical time $t_{\rm dyn}\propto R_0$, the fluence is then $\propto t_{p\gamma}^{-1}t_{\rm dyn}\propto R_0^{-1}$.
On the other hand, since $B_0\propto R_0^{-1}$, $E_{\pi , \rm damp}^{\rm ob} \propto R_0$, leading to a spectral turnover at higher energy for larger $R_0$, i.e., the neutrino spectrum extends to higher energy. 

The decrease of spectral peak somewhat compensate with the increase of the peak energy (corresponding the muon or pion damping energy for $\nu_e$ or $\nu_\mu$, respectively), leading to comparable fluences at high energies for different $R_0$, except for the case of $R_0=10^{13}$cm, which is an optically thick case (Section \ref{sec:evolution}). 

In the optically thick case, the strong cooling of high-energy nucleons by PIs at small radii suppresses pion production at large radii, where muon and pion damping is weak. Thus, for electron neutrinos and muon neutrinos above their damping energy ($E_{\nu}^{\rm ob} \sim 10^{7} {\rm GeV}$ for muon damping and $E_{\nu}^{\rm ob} \sim 10^{8} {\rm GeV}$ for pion damping, see in Figure \ref{fig:R0}), the production is strongly suppressed. In this case, the neutrino fluence is much smaller than that with large $R_0$, showing a power law $\Phi_\nu\propto E_\nu^{-4}$, similar to the instantaneous approximation.

The impact of $\Gamma$ on neutrino spectrum is shown in Figure \ref{fig:Gamma}. It can be seen that the fluence spectrum is sensitive to $\Gamma$ at low energies but not at high energies. By Figure \ref{fig:Gamma}, one can find the fluence and the Lorentz factor follow a relation, $(E_{\nu}^{\rm ob})^2\Phi_{\nu}^{\rm ob} \propto \Gamma^{-4}$, in the low-energy range where the contribution of neutrinos by resonant channel dominates. As $B_0 \propto \Gamma^{-1}$, the pion damping energy should be $E_{\pi,\rm damp}\propto\Gamma$ if synchrotron cooling dominates adiabatic cooling. The fact that the fluence decreases but the pion damping energy increases with $\Gamma$ results in that the neutrino fluence at high energies is very insensitive to $\Gamma$.

\subsection{Neutrino flavor fraction}\label{subsec:3.2}


As known, for neutrinos from the chain decay of charged pions, the neutrino flavor fractions are $(f_{e,S}:f_{\mu,S}:f_{\tau,S}) = (1/3:2/3:0)$ (Here $f_{\alpha} \equiv \Phi_{\nu_{\alpha}} / \Phi_{\nu_{\rm all}}$, $\alpha=e,\mu,\tau$; the subscript "S" denotes the flavor fraction in sources). However, if suffering strong muon damping, they become $(0:1:0)$. In the instantaneous approximation, the flavor ratio transition happens sharply at energy corresponding muon damping, but in our careful consideration of time evolving spectrum, the case is different.
\begin{figure}[t]
    \centering
    \includegraphics[width=1\linewidth]{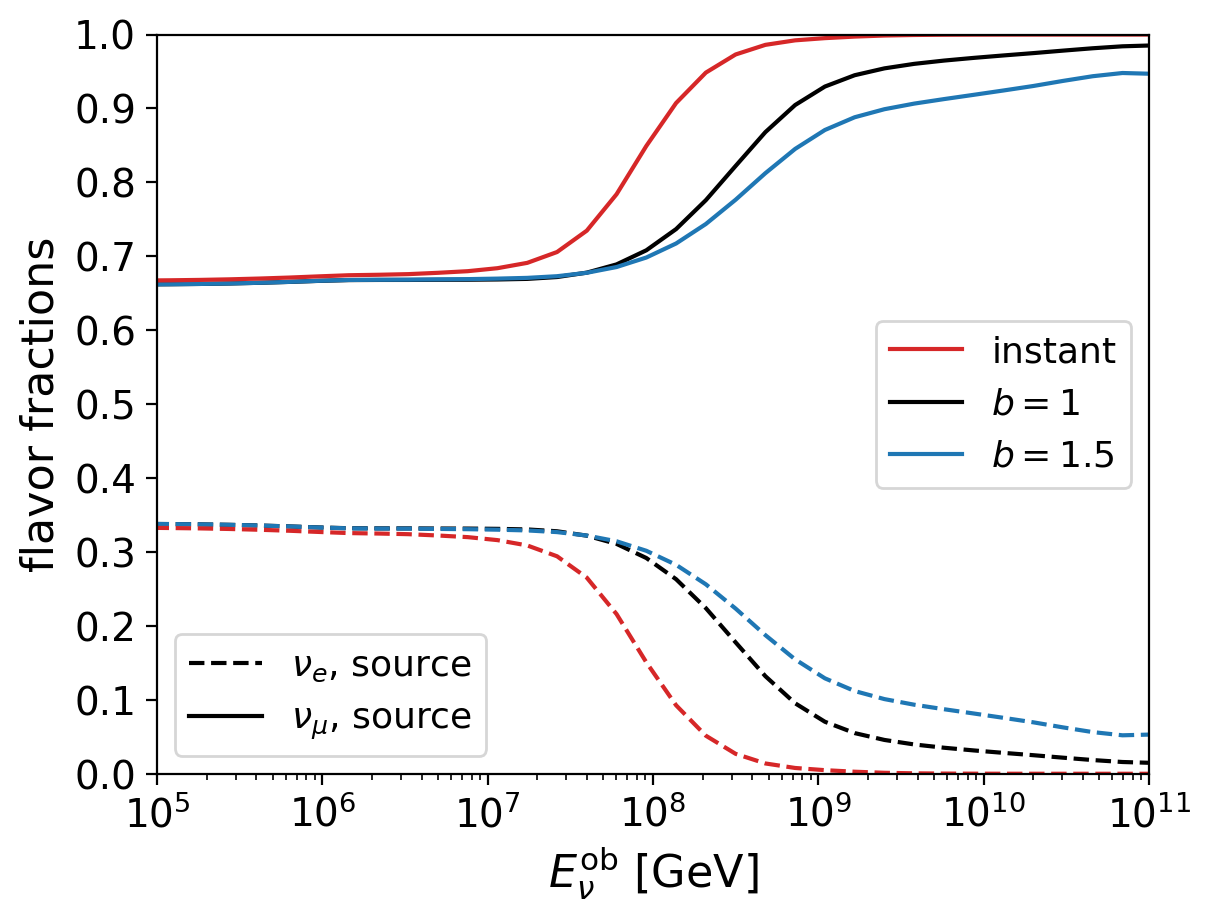}
    \includegraphics[width=1\linewidth]{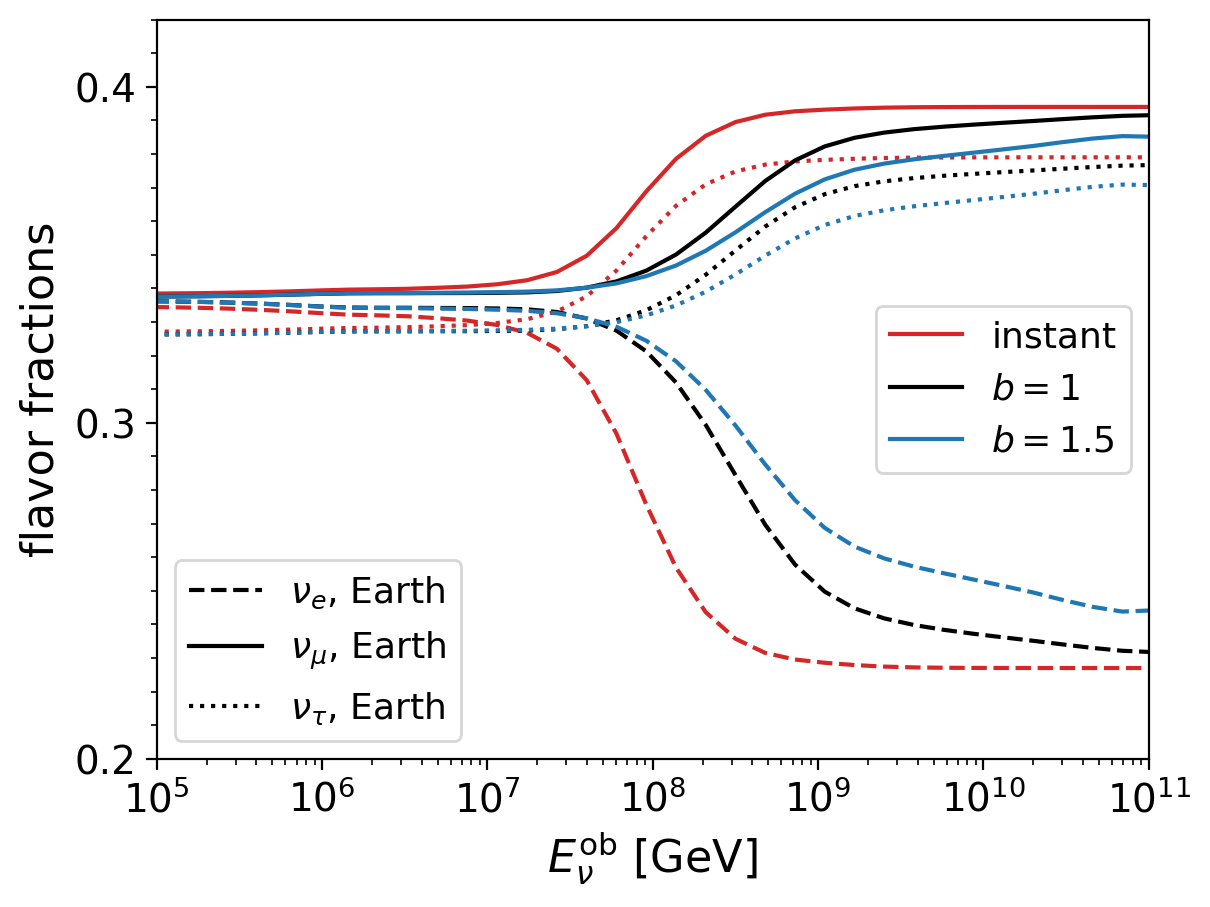}
    \caption{Neutrino flavor fraction as function of neutrino energy. The upper and lower panels show the flavor fractions in the source and at the Earth, respectively. The case of instantaneous approximation (red lines) is shown for comparison.}
    \label{fig:flavor}
\end{figure}


As shown in Figure \ref{fig:flavor}, the transition of neutrino flavor fractions caused by muon damping occurs roughly at $E_\nu^{\rm ob}\sim 10^{8}$GeV, somewhat larger than in the instantaneous approximation, $E_\nu^{\rm ob}\sim E_{\mu, \rm damp}^{\rm ob}(R_0)/3 \sim 10^{7.5}$GeV. This is because in the time-integrated model, the production of electron neutrinos at large radii, where muon damping is not severe, is still significant, as Figure \ref{fig:magnetic_field_decay} shows. Then the change of flavor fractions for the time-integrated model is slower than that of instantaneous approximation.

There should be another critical neutrino energy that is related to pion damping. Because pion damping suppresses the muon neutrino fluence, there is a break at $E_\nu^{\rm ob}\sim E_{\pi, \rm damp}^{\rm ob}(R_0)/4 \sim 10^{9} {\rm GeV}$. The decrease of electron neutrino fraction become slower for larger energy. In Figure \ref{fig:flavor}, this break is significant in our time-integrated models due to the non-negligible electron neutrino fraction. However, in the instantaneous approximation, the electron neutrino fraction is almost zero at the pion damping energy, thus the second critical point does not appear.

Neutrinos undergo oscillation during propagation. We show the flavor fraction when neutrinos arrive at the Earth in the lower panel of Figure \ref{fig:flavor}. At arrival, the flavor fraction in sources is transformed into $f_{\alpha, \oplus} = \Sigma_{\beta} P_{\beta \alpha} f_{\beta,S}$, where $P_{\beta \alpha} = \Sigma_i \lvert U_{\beta i} \rvert \lvert U_{\alpha i} \rvert$ is the average flavor transition probability, and $U$ is the lepton mixing matrix. We adopt the best-fit values of the neutrino mixing parameters from NuFit 5.3 \citep[][]{2020JHEP...09..178E,Nufit} in the calculation. We see that the flavor fractions at the Earth change from  $(f_{e,\oplus}, f_{\mu,\oplus}, f_{\tau,\oplus}) \approx (0.33 : 0.34 : 0.33)$ to $(0.23 : 0.40 : 0.38)$ due to the muon damping. Tau neutrinos appear because of neutrino oscillation.

Again, we emphasize two features of flavor fraction in contrast to that of instantaneous approximation: the transition occurs at relatively larger neutrino energy; and the electron (muon and tau) neutrino fraction does not decrease (increase) to the level of the instantaneous approximation. These are all due to the neutrino production at radius much larger than $R_0$.

\section{Conclusion and discussion} \label{sec:discussion}

In this work, we consider time-evolving and then time-integrated neutrino production in GRB jets during the whole period of the expansion of jets.  We solve continuity equations for nucleons, pions and muons to account for particle production, cooling and decay in the evolution. We use model \textit{Sim-B} from H10 for numerical calculation of pion production.
We find that the jet expansion affects pion damping effect, and result in significantly different neutrino spectra for very high energy neutrinos compared with that of instantaneous approximation. Under the fiducial parameter setting, our neutrino fluence is significant higher than that of instantaneous approximation for neutrinos of $E_{\nu}^{\rm ob} \gtrsim 1$ EeV; the faster decrease of magnetic field leads to larger neutrino fluence due to contribution from large radii (see Figure \ref{fig:magnetic_field_decay}).
Moreover, the jet expansion also affect the muon damping effect, and results in different neutrino flavor fraction as function of neutrino energy (Figure \ref{fig:flavor}), compared to that of the instantaneous approximation.
Finally, we also find that the neutrino fluence at very high energies, say, $\ga10$ EeV, is not sensitive to the dissipation radius $R_0$ and  the jet Lorentz factor $\Gamma$, except for the optically thick case (Figures \ref{fig:R0} and \ref{fig:Gamma}).

The dissipation radius, $R_0$, is a crucial physical parameter in GRB models, which also affect much the neutrino spectrum \citep[Figure \ref{fig:R0}; see also][]{2013PhRvL.110l1101Z,Ou:2024bna}. The adopted fiducial value $R_0=10^{14}$cm may correspond to the typical case of internal shock model \citep[e.g.,][]{1994ApJ...430L..93R}, whereas $R_0=10^{16}$cm to the magnetic dominated model \citep[e.g.,][]{2003astro.ph.12347L,2011ApJ...726...90Z}, and $R_0=10^{13}$cm to the dissipative photosphere model \citep{2005ApJ...628..847R}. As discussed in Section \ref{sec:nu spectrum}, the fluence is roughly $\propto R_0$ for PeV neutrinos. The pion damping energy increases with $R_0$, resulting in similar neutrino fluence for internal-shock and magnetic-dominated models at $E_{\nu}^{\rm ob} \ga 10^{10}$GeV. In dissipative photosphere model, the nucleon cool significantly in small radii and high energy neutrino production is suppressed. Thus, high-energy neutrino emission can be a probe for GRB prompt emission model. 
However, note also that in the framework of internal shocks, there should be residual collisions in larger radii generating X-ray or optical prompt emission \citep{residual_collision}. These additional photons from residual collisions may enhance the neutrino production at high energies, say, EeV-range, by PIs at large radii. 

We should also note that some previous works on GRB neutrino emission have considered energy dissipation at a wide range of radii and the contribution to UHE neutrinos \citep[e.g.,][]{2006PhRvD..73f3002M,2008PhRvD..78b3005M,2015NatCo...6.6783B,2017ApJ...837...33B}. These works show that large dissipation radii have effects on the peak neutrino energy, and generate high-energy neutrino spectrum that is flatter than $E_\nu^{-4}$, expected in the one-zone model. 


Non-detection of GRB neutrinos has put stringent constraints on GRB physical parameters. For the fiducial parameter values, the diffusive neutrino intensity at $\sim0.1$~PeV from GRBs can be estimate by $I_{\nu,{\rm GRB}}\approx (c/4\pi)f_z \dot{n}_{\rm GRB} t_{\rm H} (E_{\nu}^{\rm ob})^2N_{\nu}^{\rm ob} \sim 3\times 10^{-11} {\rm GeV \, cm^{-2} s^{-1} sr^{-1}}$, with the local GRB event rate density $\dot{n}_{\rm GRB} \sim 1 {\rm Gpc^{-3} yr^{-1}}$ \citep[e.g.,][]{2010MNRAS.406.1944W}, $f_z\sim3$ accounts for the redshit distribution of GRB event rate density \citep{1998PhRvD..59b3002W}, and $t_{\rm H}\sim10$Gyr the universe age. This is smaller than $1\%$ of the diffuse neutrino flux, satisfying the constraint of GRB neutrinos by IceCube observation \citep{ICGRB2022}. 

EeV neutrinos are largely contributed by the multi-pion channel, which results in the neutrino spectrum increases slowly with energy above the break $\sim 1$PeV. In the case of fiducial parameter values, the fluence at the spectral peak, around 0.1EeV, is roughly $\sim 1 {\rm GeV \, cm^{-2}}$. We can estimate the detection rate of GRBs by the next generation neutrino telescopes, GRAND200k and IceCube-Gen2, which are sensitive for UHE neutrinos. For example, GRAND200k's instantaneous sensitivity is $\sim0.1{\rm GeV \, cm^{-2}}$ for EeV neutrinos at zenith angle $\theta_z = 90^{\circ}$ \citep[See Figure8 in][]{GRAND_whitepaper}, thus GRBs are detectable up to 300 Mpc, and the GRB detection rate is $\sim 0.1(\Delta\Omega/4\pi) {\rm yr}^{-1}$, where $\Delta\Omega$ is the field of view of telescope. 

Adiabatic cooling is considered for all charged particles in the calculation in this work. However, the particles with large enough energy may not be confined by the magnetic field of the plasma, e.g., the Larmor radius of protons is $r_L\propto E_p/B \propto E_p R^{b}$, increasing with energy and radius, and may be larger than the size of the plasma for large $E_p$ or $R$. These protons can propagate free of adiabatic cooling, and contribute more to the neutrino production at large radius, leading to larger neutrino fluence above damping energy than the results in this work. Removal of adiabatic cooling will lead to larger damping energy of muons or pions, especially when adiabatic cooling dominate synchrotron cooling (e.g., the case of Figure \ref{fig:muon}). If protons did escape, our results can be considered as lower limits to neutrino emission.

We only consider the main neutrino production channel, pion chain decay, in our calculation, but there are other channels in the PIs, e.g., neutron decay and kaon decay \citep[][and references therein]{kaon_neutron}. Neutron decay mainly occurs outside of the source due to the long decay time of neutrons, much longer than any other timescales concerned in the process. The energy of electron neutrino from neutron decay is  small, i.e., about a fraction $(m_n - m_p) / m_n \sim 10^{-3}$ of the primary neutron, much smaller than that from pion decay and negligible. 
Kaons decay is the second most important channel for high energy neutrino production. Due to their heavy mass and short decay time, the damping energy of kaons is 30 times larger than that for pions. Thus, kaons decay can contribute some electron neutrinos to the high energy region suffering muon damping.

\begin{acknowledgments}
The authors thank Tianqi Huang and Xishui Tian for helpful discussions. This work is supported by the Natural Science Foundation of China (No. 11773003, U1931201) and the China Manned Space Project (CMS-CSST-2021-B11).
\end{acknowledgments}

%






\appendix
\section{Simplified model of photopion interactions and nucleon cooling}
\label{sec:pion production}
In this work we follow the simplified model B (\textit{Sim-B}) in H10 to treat PIs between nucleons (protons and neutrons) and gamma-ray photons and the related nucleon cooling. Note, we do not distinguish protons and neutrons here in PIs and nucleon cooling. As defined in model \textit{Sim-B}, the PIs are split into different interaction types (ITs): Pion production is separated into resonant, direct and multi-pion channels; For each channel the interactions are split into ITs according to the photon energy ranges (in the rest frame of the nucleon), where the cross sections, the inelasticities, the multiplicities and pion production rates, etc, are treated separately with different approaches.

Consider the jet propagates to the source-frame radius $R$. In the rest frame of jet, if the nucleon number per unit energy interval is $N_{p}(E_p,R)$, and the photon number per unit energy interval per unit volume is $n_\gamma(\epsilon,R)$, the number production rate of charged pions, including $\pi^+$ and $\pi^-$, per unit energy interval at pion energy $E_\pi$ and radius $R$ can be given by, following Equation (28) in H10,
\begin{equation} \label{eq: Qpi}
    Q_{\pi}(E_{\pi},R) = 
        \sum_{\rm IT} N_{p}\left(\frac{E_{\pi}}{\chi^{\rm IT}}, R \right) \frac{m_{p}c^{3}}{E_{\pi}} \int_{\epsilon_{\rm th}/2}^{\infty} {\rm d}y\, n_{\gamma}\left(\frac{m_{p}c^{2} y \chi^{\rm IT}}{E_{\pi}}, R \right) M_{\pi}^{\rm IT} f^{\rm IT}(y),
\end{equation}
where the sum adds up all ITs, $\epsilon_{\rm th}\simeq150$ MeV is the threshold, $y\equiv E_{\pi} \epsilon / \chi^{\rm IT} m_p c^{2}$, $\chi^{\rm IT}$ is the fraction of the initial nucleon energy deposited in one pion, $M_{\pi}^{\rm IT}$ is the multiplicity (i.e., the average number of pions that are produced), and $M_{\pi}^{\rm IT} f^{\rm IT}$ is the production rate of pions with specific function $f^{\rm IT}$ relevant to the cross section. 

As protons and neutrons are not distinguished, the PIs are simplified to be a cooling process. The cooling rate of a nucleon, proton or neutron, with energy $E_p$ due to PIs can be calculated following Appendix B in H10. Given the target photon spectrum $n_\gamma(\epsilon,R)$, the cooling rate of a nucleon at $R$ is
\begin{equation}
    t_{p\gamma}^{-1}(E_{p},R) = \sum_{\rm IT} \Gamma^{\rm IT}(E_{p},R) K^{\rm IT},
\end{equation}
where
\begin{equation}
    \Gamma^{\rm IT}(E_{p},R) = \int_{\epsilon_{\rm th} m_{p} c^{2}/2 E_{p}}^{\infty} {\rm d}\epsilon \, n_{\gamma}(\epsilon,R)f^{\rm IT} \left(\frac{E_{p} \epsilon}{m_{p} c^{2}}\right)
\end{equation}
is the interaction rate of a nucleon, and $K^{\rm IT}$ is the inelasticity.

For each interaction type IT, the definition of the energy range, and the specific values of $\chi^{\rm IT}$, $M_{\pi}^{\rm IT}$, $K^{\rm IT}$ and the cross section can be found in Tables 4, 5 and 6 in H10 for resonant, direct and multi-pion channels, respectively, and function $f^{\rm IT}$ in Equations (31) and (32) therein for resonances, (33) for direct channel, and (40) for multi-pion channel.

\section{Secondary spectrum from weak decays}
\label{sec:decay secondary distribution}
For a decay chain $a\rightarrow b$, the probability distribution function of secondary particle $b$, i.e., the normalized particle number per unit energy interval, can be described by $p_{a \rightarrow b}(E_b;E_a)=F_{a \rightarrow b}(E_{b}/E_{a})/E_{a}$, where $F_{a \rightarrow b}(r)$  is a dimensionless function,  with $r=E_{b}/E_{a}$. We adopt the forms of $F_{a \rightarrow b}(r)$ following \cite{Lipari2007}. For secondary muons and muon neutrinos from pion decays, we take
\begin{equation}\label{eq:Fpi2mu}
    F_{\pi \rightarrow \mu}(r) = \frac{1}{1-r_{\pi}} \Theta(r-r_{\pi}),
\end{equation}
\begin{equation}\label{eq:Fpi2numu}
    F_{\pi \rightarrow \nu_{\mu}}(r) = \frac{1}{1-r_{\pi}} \Theta(1 - r_{\pi} - r),
\end{equation}
where $r_{\pi} = (m_{\mu}/m_{\pi})^{2}$. For secondary electron and muon neutrinos from muon decays, we take
\begin{equation}\label{eq:Fmu2numu}
    F_{\mu \rightarrow \nu_{\mu}}(r) = \frac{5}{3} - 3 r^2 + \frac{4}{3} r^3,
\end{equation}
\begin{equation}\label{eq:Fmu2nue}
    F_{\mu \rightarrow \nu_{e}}(r) = 2 - 6 r^2 + 4 r^3.
\end{equation}
Note, because neutrinos and anti-neutrinos are summed up in this work, we has adopted the functions for unpolarized muons. 

\section{Instantaneous approximation for neutrino production}
\label{sec:instan approx}
In the usual approaches for the calculation of neutrino production in GRB jets, only contribution within one dynamical time around $R_0$ is considered, not caring about the evolution and time-integral of the neutrino spectrum during jet expansion. In these kinds of approaches, either steady-state solution for pion and muon spectra or time-dependent evolution within only one dynamical time is considered \citep[e.g.,][]{2006PhRvD..73f3002M,HummerSimB,2012PhRvL.108w1101H,2015NatCo...6.6783B,2017ApJ...837...33B}. We would like to call these approaches the "instantaneous approximation" here. 

To calculate the neutrino spectrum in the instantaneous approximation, the following approach is adopted. First, to obtain the steady-state solutions for pion and muon spectra at $R=R_0$, respectively, we take 
\begin{equation}\label{eq:steady-meson}
    N_x(E_x,R_0) = Q_x(E_x,R_0) {\rm min}[\tau_{x}(E_x),t_{x,{\rm cool}}(E_x,R_0)],
\end{equation}
where $x=\pi$ or $\mu$. Note, no adiabatic cooling is considered. The production rates, $Q_x$'s, for pions or muons are calculated by the same way described in Appendix \ref{sec:pion production} and \ref{sec:decay secondary distribution}. Next, the neutrino production rate $Q_\nu(E_\nu,R_0)$ is given by Equation (\ref{eq:Qmu,Qnu}) taking the steady-state spectra of pions and muons, $N_x(E_x,R_0)$, in Equation (\ref{eq:steady-meson}). Finally, the total neutrino spectrum is calculated as multified by the dynamical time $t_{\rm dyn}(R_0)=t_0$,
\begin{equation}
    N_{\nu}(E_\nu) = Q_{\nu}(E_\nu,R_0) t_{0}.
\end{equation} 
So in such approximation, the instantaneous neutrino production rate around $R_{0}$ is used, no adiabatic cooling is considered, and the contribution from large radii is ignored.




\bibliography{sample631}{}
\bibliographystyle{aasjournal}



\end{document}